\documentclass[12pt,journal,draftclsnofoot,onecolumn]{IEEEtran}

\usepackage{color}
\usepackage{bm}
\usepackage[pdftex]{graphicx}
\DeclareGraphicsExtensions{.pdf,.jpeg,.png}
\usepackage{epstopdf}
\usepackage{tikz}
\usetikzlibrary{patterns}
\usepackage{pgfplots}
\usepackage{amsmath}
\usepackage{dsfont}
\allowdisplaybreaks[4]

\usepackage{cite}

\ifCLASSINFOpdf
\else
\fi
\usepackage{amsmath}
\usepackage{algorithmic}

\usepackage{array}

\ifCLASSOPTIONcompsoc
  \usepackage[caption=false,font=normalsize,labelfont=sf,textfont=sf]{subfig}
\else
  \usepackage[caption=false,font=footnotesize]{subfig}
\fi
\usepackage{url}
\usepackage{mathalfa}
\usepackage{amsfonts}
\usepackage{color}
\usepackage{makecell}
\newtheorem{theorem}{Theorem}
\newtheorem{lemma}{Lemma}

\graphicspath{{fig/}}
\usepackage{diagbox}
\usepackage{booktabs}
\usepackage[bottom]{footmisc}

\begin{document}

%
\title{User-Centric Secure Cell Formation for Visible Light Networks with Statistical Delay Guarantees}
%
%
%
\author{Lei~Qian,~\IEEEmembership{Student~Member,~IEEE,}
        Xuefen~Chi,~
        Linlin~Zhao,~\IEEEmembership{Member,~IEEE,}
        Mohanad~Obeed,~
        and~Anas~Chaaban,~\IEEEmembership{Senior~Member,~IEEE}
\thanks{Manuscript received XX XX, 2020; revised XX XX, 2020. The work of L. Qian, X. Chi and L. Zhao was supported in part by National Natural Science Foundation of China under Grant 61801191, in part by Jilin Scientific and Technological Development Program under Grant 20180101040JC and 20200401147GX, in part by the China Scholarship Council under Grant 201906170201. The work of M. Obeed and A. Chaaban was supported by the Natural Sciences and Engineering Research Council of Canada (NSERC).}
\thanks{L. Qian, X. Chi and L. Zhao are with the College of Communication Engineering, Jilin University, Changchun, China, email: (qianlei16@mails.jlu.edu.cn, chixf@jlu.edu.cn, zhaoll13@mails.jlu.edu.cn).}
\thanks{M. Obeed and A. Chaaban are with the School of Engineering, University of British
Columbia, Kelowna, BC, V1V 1V7, CA, email: (anas.chaaban@ubc.ca, mohanad.obeed@ubc.ca).}}
\IEEEaftertitletext{\vspace{-1.5\baselineskip}}
\maketitle
\begin{abstract}
In next-generation wireless networks, providing secure transmission and delay guarantees are two critical goals. However, either of them requires a concession on the transmission rate. In this paper, we consider a visible light network consisting of multiple access points and multiple users. Our first objective is to mathematically evaluate the achievable rate under constraints on delay and security. The second objective is to provide a cell formation with customized statistical delay and security guarantees for each user. First, we propose a user-centric design called secure cell formation, in which artificial noise is considered, and flexible user scheduling is determined. Then, based on the effective capacity theory, we derive the statistical-delay-constrained secrecy rate and formulate the cell formation problem as a stochastic optimization problem (OP). Further, based on the Lyapunov optimization theory, we transform the stochastic OP into a series of evolutionary per-slot drift-plus-penalty OPs. Finally, a modified particle swarm optimization algorithm and an interference graph-based user-centric scheduling algorithm are proposed to solve the OPs. We obtain a dynamic independent set of scheduled users as well as secure cell formation parameters. Simulation results show that the proposed algorithm can achieve a better delay-constrained secrecy rate than the existing cell formation approaches.
\end{abstract}


%
\IEEEpeerreviewmaketitle

\section{Introduction}
%
%
%
%
\IEEEPARstart{S}{ince} the radio frequency (RF) spectrum has become congested, visible light communication (VLC) is promoted as one of promising alternatives for future wireless networks \cite{VLC+6G}. The achievable rate of VLC can easily reach several Gbps owing to its unlicensed broadband visible light spectrum (400 to 800 THz), and the advanced modulation and coding techniques  \cite{OB-2013-Gbps}.

Protecting personal information from eavesdropping has always been a big concern in wireless networks. Due to the fact that light can be blocked by walls or by opaque objects, indoor VLC systems are considered more secure compared to traditional RF networks. However, user's confidential information may still be intercepted by other users when VLC systems are deployed in public areas, such as shopping malls, libraries, hospitals, airports, etc. Due to this, physical layer security (PLS) has been introduced into VLC systems to improve security by exploiting the inherent randomness of the physical medium and the difference between legitimate channels and wiretap channels \cite{Southeast-2018-TCOM}-\hspace{1sp}\cite{Oxford-2018-TWC}. 
Considering the illumination constraint and the physical characteristics of a light source, the authors of \cite{Southeast-2018-TCOM} derived the secrecy capacity of VLC to characterize the maximum achievable secrecy rate for scenarios with one access point (AP) and one eavesdropper. 
For the multiple APs and one eavesdropper scenario, lower and upper bounds of secrecy capacity were derived for spatial modulation based VLC in \cite{VLCPLS-spatialmodu}. For single-user multiple-input single-output (MISO) VLC systems, different PLS techniques were proposed to improve the secrecy performance, including zero-forcing \cite{Lampe-2015-JSAC}, friendly jamming \cite{Lampe-2014-globecom}, robust beamforming \cite{Mashuai} and artificial noise \cite{Anas-2019-TCOM}. Recently, based on deep reinforcement learning techniques, a smart beamforming approach was proposed for the non-ideal wiretap channel in \cite{VLCPLS-deepRL}. For a simultaneous lightwave information and power transfer system, the tradeoff between the secrecy capacity and the harvested energy was investigated in \cite{VLCPLS-harvest}, and a protected zone-based secure scheme was investigated in \cite{VLCPLS-SLIPT}.
For a VLC system consisting of one AP and multiple users (one legitimate user and several eavesdroppers), secrecy outage probability was investigated in \cite{Oxford-2018-WCL} under the assumption that the eavesdroppers are randomly distributed. For more practical VLC networks, consisting of multiple APs and multiple users, the ergodic secrecy rate was derived and an eavesdropper-free disk-shaped protected zone is proposed in \cite{Haas-2018-JSAC}. Besides, a closed-form secrecy outage probability was derived and a light-emitting diode (LED) selection approach was proposed in \cite{Oxford-2018-TWC}. However, both \cite{Haas-2018-JSAC} and \cite{Oxford-2018-TWC} selected only one AP to serve the legitimate user, which greatly restricts the transmission rate and does not make efficient use of LED resources. 

One effective way to efficiently utilize the LED resources is to carefully form the cells in VLC networks, and this issue attracts more attention in the literature \cite{Haas-2016-attocell}-\hspace{1sp}\cite{UCQoE-2019-XuBao}. 
Inheriting from traditional RF networks, some works proposed the network-centric VLC cell formation.
In the network-centric design, the cells are formed by firstly grouping the APs that are close to each other, then associating the users to these APs' groups. 
Since the coverage area of a VLC AP is considered much smaller than the coverage area of a RF AP (due to the link blockages, the users' field of view (FoV), and the high attenuation of VLC links with distance), the edge-users in the overlapping coverage area of network-centric VLC cells may suffer from serious inter-cell interference \cite{interference-attocell-2018-JLT}.
Recently, the emerging user-centric principle is adopted in VLC cell formation, which constructs amorphous cells dynamically by considering the traffic during the current timeslot \cite{XuanLi-2016-TWC}-\hspace{1sp}\cite{UCQoE-2019-XuBao}.
In the user-centric design, each user selects a subset of APs that are the most qualified to serve that user, where these subsets may overlap with each other.
Therefore, a user-centric design is more capable of providing customized service for each user, because the cells can be adjusted adaptively according to users' individual requirements \cite{XuanLi-2018-survey}. For example, triggered by users' mobility, handover is an inevitable problem in network-centric design, while in the user-centric design, cells shape can be updated instantly by tracking the users' real-time locations \cite{RZhang-2018-TWC}. Moreover, user scheduling, cell formation, and resource allocation procedures can be implemented jointly in the user-centric design. Therefore, with more flexibility, user-centric cell formation is promising to guarantee the customized secrecy transmission requirement for each user. However, up to our best knowledge, this is the first paper that investigates and optimizes the secure user-centric design for VLC networks.

In addition to secrecy, users' various delay quality-of-service (QoS) demands should also be guaranteed, particularly for delay-sensitive traffic \cite{2019-ESR}. Previous works \cite{Southeast-2018-TCOM}-\hspace{1sp}\cite{Oxford-2018-TWC} 
analysing secrecy performance for VLC systems did not consider the legitimate users' delay requirements. Other works investigated delay guarantees in user-centric VLC networks without taking users' security requirements into account\cite{RZhang-2018-TWC}\cite{Chi-2018-COPP} \cite{XuanLi-2016-Globecom}.
 A user-centric delay-aware association algorithm is proposed in \cite{RZhang-2018-TWC} based on the largest weighted queue scheduling. To mathematically quantify the delay QoS from a statistical perspective, \cite{Chi-2018-COPP} and \cite{XuanLi-2016-Globecom} introduced the effective bandwidth (EB) and effective capacity (EC) theories into user-centric VLC cell formation. Emerging and booming in the last decade, EB and EC theories provide a concise and powerful mathematical framework to investigate the performance of wireless networks under certain statistical delay constraints. Based on the queuing model, EB indicates the constant service rate needed by a certain arrival process under a statistical delay limit \cite{EB-1995}. As the dual of EB, EC represents the constant arrival rate that can be supported with a certain service process subject to a statistical delay constraint \cite{Wu-2003}. In \cite{Chi-2018-COPP}, EB is utilized to modify the proportional fairness priority factor in user-centric multi-user scheduling. In addition, EC is treated as the weight of each link between users and APs in a user-centric VLC cell formation problem \cite{XuanLi-2016-Globecom}. However, the authors of \cite{XuanLi-2016-Globecom} did not consider the impact of cell formation on the service process in their derivation of EC.

Note that, on one hand, once EC is employed to develop a delay-driven scheduling (or cell formation) algorithm, no matter in VLC or RF networks, a closed-form EC function is needed for scheduling scheme design. On the other hand, since EC characterizes delay based on the random queuing model, deriving EC requires a mathematical expression of the stochastic service process, which strongly depends on the scheduling scheme. Therefore, EC derivation and scheduling scheme design are two tightly coupled problems, an aspect which has not been considered in the previous literature to the best of our knowledge. An EC expression ignoring the impact of scheduling scheme is not convincing, which would result in a designed scheduling algorithm that fails to provide delay QoS guarantees. Besides, a quantified statistical delay requirement is depicted by the delay violation probability.
It is worth investigating how to handle a probabilistic delay guarantee goal in a practical network. More specifically, the probabilistic delay guarantee goal is a theoretical metric on a large time scale, while the operations in practical network are updated dynamically on a small time scale. Therefore, the main challenge is how to provide long-term statistical delay guarantees with secrecy while adjusting cell formation over a short time scale. 
Similar dynamic wireless resource control problems can be formulated as a Markov Decision Process (MDP) problem, and can be solved by the Bellman equation and value iteration methods \cite{MDP-sensor}. However, the computational complexity of the MDP-based approaches may cause curse of dimensionality due to the large volume of the state and action space. To reduce the complexity, approximate MDP and stochastic learning methods are proposed in delay-aware resource control \cite{2012-TIT-delay-aware-control}. Besides, without relying on the state transition, the Lyapunov optimization theory provides an effective framework to solve this kind of wireless resource control problem with relatively low complexity \cite{Neely-book}, which is used as the main tool of this paper.

In this paper, a multi-AP multi-user VLC network is considered, in which any user is a potential eavesdropper to other users in the same room. We propose a dynamic user-centric cell formation algorithm with customized heterogeneous statistical delay and security guarantees. The main contributions of this paper can be summarized as follows.

\begin{itemize}
\item Jointly considering the user scheduling and PLS, we propose a user-centric cross-layer design called secure cell formation. In the secure cell of each user, a specific level of artificial noise is adopted to enhance the transmission security. From an information theoretic perspective, we derive the achievable secrecy rate for each user as a function of the secure cell formation parameters.
\item Based on the EC theory, we derive the effective secrecy rate (ESR) to characterize the delay-constrained achievable secrecy rate for each user. With the help of EB, we construct long-term statistical delay guarantee inequalities, which contribute to depicting the heterogeneous probabilistic delay requirements. Then, we formulate the user-centric secure cell formation with multiple long-term delay guarantee constraints as a stochastic optimization problem.
\item Since the probabilistic delay guarantee goal is a theoretical metric, we bridge the theoretical metric and practical network control with the help of Lyapunov optimization theory \cite{Neely-book}. Lyapunov virtual queues are employed to evaluate how much service is needed for each user in future timeslots to achieve the long-term statistical delay guarantee goals. In other words, we decompose the long-term delay guarantee goals into gradually changed short-term targets for each timeslot. According to these short-term targets, we transform the stochastic optimization problem into several temporal evolutionary per-slot deterministic optimization problems with the help of the drift-plus-penalty (DPP) method \cite{Neely-book}.
\item Since the per-slot optimization problem is a mixed high-dimensional integer programming problem, we first decouple it into intra-cell secure parameters optimization sub-problem and user-centric scheduling sub-problem, and then propose the modified particle swarm optimization (PSO) algorithm \cite{PSO1999} and interference graph (IG) based greedy scheduling algorithm to tackle the sub-problems, respectively. Solving the deterministic per-slot optimization problem over timeslots, we can dynamically adjust the secure transmission rate according to user's own statistical delay requirement by updating the secure cell formation. 
\item In the numerical examples, we evaluate the performance of the proposed secure cell formation algorithm on both short-term and long-term. In addition, we also analyse the effects of diverse VLC characteristics on the ESR, such as FoV, users' density, etc. Simulation results demonstrate that the proposed cell formation algorithm is capable of providing a higher delay-constrained secrecy rate than the existing user-centric cell formation methods.

\end{itemize}

The rest of this paper is organized as follows. The system model is introduced in Sec. \ref{section-system model}. The artificial noise aided ESR are derived in Sec. \ref{section-pri-rate}. In Sec. \ref{section-Lya}, we present the problem formulation and the proposed Lyapunov optimization solution. Simulation results are presented and discussed in Sec. \ref{section-simulation}. Finally, the paper is concluded in Sec. \ref{section-conclusion}.
 
\section{System Model}
\label{section-system model}

\subsection{System description}
\label{section-system-description}
\begin{figure}[!t]
\centering
\includegraphics[width=4.5in]{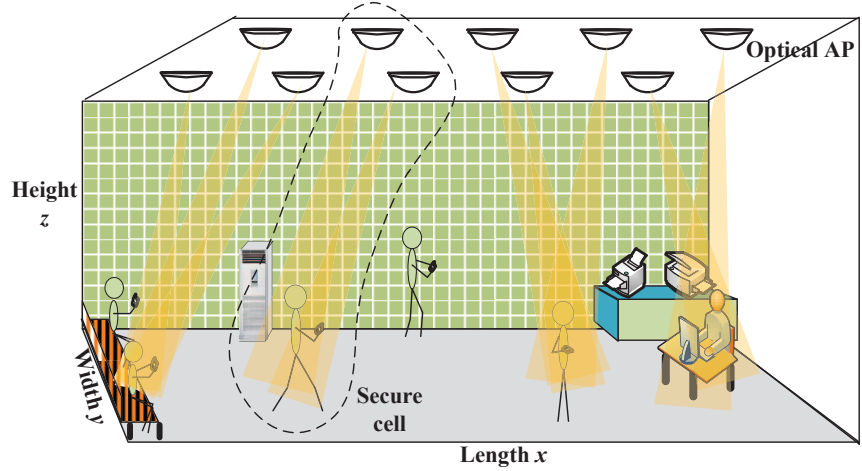}%
\caption{Indoor multi-AP multi-user VLC system with secure cells.}
\label{Fig-scenario}
\end{figure}

In this paper, an indoor downlink private VLC system is considered, as shown in Fig. \ref{Fig-scenario}. The room size is $x \times y \times z$ (length $\times$ width $\times$ height). The system contains $K_a$ APs and $K_u$ users. Each AP relies on an LED array constituted by $m$ illumination LEDs. The set of users is denoted by $\Phi  = \left\{ 1,\ldots,{K_u} \right\}$. For any communicating user, all the other users in the same room are considered as malicious eavesdroppers. For user $j \in \Phi$, the set of eavesdroppers can be expressed as ${\Psi _j} = \left\{ {k \in \Phi \left| {\;k \ne j} \right.} \right\}$.

Each user has its own statistical delay QoS requirement, consisting of a delay bound (maximum tolerable delay) and a delay bound violation probability. From a queueing theory perspective, for user $j$, it is required that $\Pr \{ {D_j} \ge D_j^{\max }\}  \approx {e^{ - {\theta _j}{\mu _j} D_j^{\max }}}$, where $D_j$ and $D_j^{\max }$ are the steady state delay and the delay bound of the user $j$, $\theta _j$ is the QoS exponent of user $j$, ${\mu_j}$ is a fixed rate determined jointly by the arrival process and the service process \cite{Wu-2003}. It is apparent that the QoS exponent $\theta _j$ plays an important role in the statistical delay QoS guarantee problem. For a given $D_j^{\max }$, a larger QoS exponent $\theta _j$ corresponds to a smaller delay violation probability, i.e., a stricter delay QoS constraint. In contrast, a smaller QoS exponent $\theta _j$ corresponds to a bigger delay violation probability, which means a looser delay QoS requirement. Hence, we use the QoS exponent vector to characterize the heterogeneous statistical delay QoS requirements of user set $\Phi$, i.e. $\bm{\theta }{\rm{ = [}}{\theta _1}, \ldots ,{\theta _{{K_u}}}{\rm{]}}$. 

To guarantee secrecy and a specific delay requirement of each user, a user-centric design called dynamic secure cell formation is proposed for the indoor downlink VLC system. Each secure cell, consisting of multiple APs and one user, sends information to its user while keeping it confidential from all other users. The time of the VLC system is divided into timeslots and we adjust cell formation at the beginning of each timeslot. 
Let ${\bf{\Pi }}(t)$ be a ${K_a} \times {K_u}$ matrix that denotes the secure cell formation matrix of the multi-user multi-AP VLC system in timeslot $t$. 
The $j$th column of ${\bf{\Pi }}(t)$, i.e. ${\bf{\Pi }}_j(t)$, is defined to characterize the connectivity between user $j$ and APs. 
The element of ${\bf{\Pi }}_j(t)$ is expressed as ${\pi _{i,j}}(t) \in \{ 0,1\}$, where ${\pi_{i,j}}(t)=1$ indicates that user $j$ is connected to AP $i$ in timeslot $t$ and ${\pi_{i,j}}(t)=0$ otherwise. 
In this paper, ${\pi_{i,j}}(t)$ depends on the user scheduling in timeslot $t$ and the set of APs who are capable to serve user $j$, which is characterized by ${\Omega _j}$. 
In particular, if user $j$ is scheduled in timeslot $t$, all of APs in ${\Omega _j}$ are allocated to user $j$, i.e., ${\pi_{i,j}}(t)=1 $ for all $i \in {\Omega _j}$. 
Considering the user's location and the FoV, the capable AP set of user $j$ is expressed as 
\begin{equation}
\label{capable AP set}
{\Omega _j}=\left\{ {i \in \{1, \ldots ,K_a \}} | (h_{i,j}^{\rm{LoS}}+\int_{{\rm{walls}}} {1\;dh_{i,j}^{\rm{NLoS}}} )>\varepsilon \right\},
\end{equation}
where $h_{i,j}^{\rm{LoS}}$ and $h_{i,j}^{\rm{NLoS}}$ represent the VLC channel gain of line-of-sight (LoS) and non-LoS links between AP $i$ and user $j$, respectively, and $\varepsilon$  here can be defined as the threshold of VLC channel gain, where it has an impact on the accuracy of interference characterization, and this is discussed in detail in Subsection ~\ref{subsection-IG}. 

\subsection{The VLC Channel model}
\label{subsection-channel model}
A direct current (DC)-biased intensity modulation direct detection (IM-DD) scheme is considered, in which the LED is driven by a fixed bias $I_{DC} \in \mathbb{R}_ +$, where $\mathbb{R}_ +$ represents the set of non-negative real numbers. The DC bias sets the average radiated optical power and adjusts the illumination level. The data signal $s \in \mathbb{R}$ is a zero-mean current signal superimposed on $I_{DC}$ to modulate the instantaneous optical power emitted from the LED. Considering the clipping distortion, the total current $I_{DC}+s$ should be constrained within the range $I_{DC} \pm \gamma I_{DC}$, where $\gamma \in [0,1]$ represents the modulation index \cite{Lampe-2015-JSAC}. Therefore, $s$ needs to satisfy a peak-power constraint characterized by $|s| \le A$, where $A=\gamma I_{DC}$. At the transmitter side, the total current $I_{DC}+s$ is converted into optical power $\eta(I_{DC}+s)$ and transmitted by the LED, where $\eta$ (W/A) is the current-to-light conversion efficiency. At the receiver side, the photo-detector (PD) of responsivity $\varpi$ (A/W) converts the received optical power into current. After removing the DC-offset $I_{DC}$ and amplifying by a transimpedance amplifier of gain $T$ (V/A), the received voltage signal $y$ is obtained.

In the formed secure cell of the scheduled user $j$ in timeslot $t$, there are $\left| {{\Omega _j}} \right|$ APs transmitting the signal to user $j$, where $\left| {{\Omega _j}} \right|$ is the size of the set ${\Omega _j}$. A vector transmission technique is adopted and each secure cell is modelled as a MISO system \cite{XuanLi-2016-TWC}. The received signal of user $j$ in timeslot $t$ is expressed as
\begin{equation}
\label{MISO-received-signal}
{y_j} (t)= {\bf{h}}_j^T(t){{\bf{s}}_j}(t){\rm{ + }}{n_j}(t),
\end{equation}
where ${{\bf{h}}_j}(t) \in \mathbb{R}_{\rm{ + }}^{\left| {{\Omega _j}} \right| \times {\rm{1}}}$ and ${{\bf{s}}_j}(t) \in \mathbb{R}^{\left| {{\Omega _j}} \right| \times {\rm{1}}}$ represent the channel gain vector and the zero-mean transmitted signal vector of user $j$ in its secure cell, respectively. The peak-power constraint of ${{\bf{s}}_j}(t)$ is expressed as 
\begin{equation}
\label{equ-peak power constraint}
{\left\| {{{\bf{s}}_j}(t)} \right\|_\infty } \le A,
\end{equation}
where $A \in \mathbb{R}_{\rm{ + }}$ is the peak power constraint. Besides, ${n_j}(t)\sim{{\mathcal N}(0,\sigma^2)}$ represents the VLC noise of user $j$, which contains shot noise and thermal noise. The VLC noise is regarded as additive white Gaussian noise with zero-mean and variance $\sigma^2=N_0 B$, where $N_0 \approx 10^{-22} \;\rm{A^2/Hz}$ \cite{2008-VLCnoise}. 

For user $k \in \Psi _j$, the received eavesdropped signal intended to user $j$ is expressed as
\begin{equation}
\label{received-signal-eve}
y_{j,k}^{{\rm{e}}}(t) = {\left( {{\bf{h}}_{j,k}^{{\rm{e}}}} (t)\right)^T}{{\bf{s}}_j}(t){\rm{ + }}n_{j,k}^{{\rm{e}}}(t),
\end{equation}
where ${\bf{h}}_{j,k}^{{\rm{e}}}(t)\in \mathbb{R}_{\rm{ + }}^{\left| {{\Omega _j}} \right| \times {\rm{1}}}$ represents the wiretap channel vector, from APs in set ${\Omega _j}$ to user $k$, $n_{j,k}^{\rm{e}}(t)\sim{{\mathcal N}(0,\sigma^2)}$ represents noise. 

In this paper, the channel gain includes both the LoS part and the non-line-of-sight (NLoS) part. We only consider the first reflection paths of NLoS links since the effect of other reflection paths is so small and can be neglected \cite{2004-NLOS}. Due to the existence of some opaque objects in the room, LoS links may be blocked. A random variable $\xi_j(t)$ is defined as a blockage indicator of timeslot $t$, where $\xi_j(t)=1$ indicates that user $j$ is unblocked and can receive through both the LoS and the first reflection links, and $\xi_j(t)=0$ means that user $j$ is blocked and only receives through the first reflection links. We assume that the blockage event occurs randomly and follows an independent and identically distributed (IID) Bernoulli distribution with parameter $\beta$, i.e. ${\xi _j}(t) \sim {\rm{Bern}}(\beta )$.
%
Then, the elements of ${\bf{h}}_j(t)$ can be expressed as 

\begin{equation}
\label{gain-prob-bob}
h_{i,j}(t)=\xi_j(t) h^{\rm{LoS}}_{i, j} + \int_{{\rm{walls}}} {1\;dh_{i,j}^{\rm{NLoS}}},\; i \in {\Omega _j},
\end{equation}
where $h_{i,j}^{{\rm{LoS}}}$ and $h_{i,j}^{{\rm{NLoS}}}$ denote the channel DC gain of LoS link and the first reflected link from AP $i$ to user $j$, respectively. Specifically, for a LoS link \cite{2004-NLOS},
\begin{equation}
\begin{aligned} 
\label{LOS-gain}
h_{i,j}^{{\rm{LoS}}}{\rm{ = }}{\eta{(L_a + 1)\delta \varpi {T}} \over {2\pi d_{i,j}^2}}g(\psi _{i,j}^{{\rm{LoS,in}}}){\cos ^{L_a}}(\psi _{i,j}^{{\rm{LoS,ir}}})\cos (\psi _{i,j}^{{\rm{LoS,in}}}),
\end{aligned} 
\end{equation}
where $d_{i,j}$ is the distance between the VLC AP $i$ and the user $j$, $L_a$ denotes the order of Lambertian emission, which is given by $L_a =  - \ln 2/\ln(\cos({\phi _{1/2}}{\rm{ }}))$ and $\phi_{1/2}$ is the semi-angle at half illumination of LEDs, $\delta$ is the physical area of the detector in a PD, $\varpi$ is responsivity of a PD, $\eta$ is the current-to-light conversion efficiency of
the LED, $T$ is the gain of the transimpedance amplifier, $\psi _{i,j}^{{\rm{LoS,in}}}$, and $\psi _{i,j}^{{\rm{LoS,ir}}}$ are the angle of incidence and irradiance between the VLC AP $i$ and the user $j$, respectively. Furthermore, the function $g(\cdot)$ represents the gain of an optical concentrator, which is characterized by 
\begin{equation}
\label{optical-concentrater}
g(\psi _{i,j}^{{\rm{LoS}},{\rm{in}}}) = \left\{ \begin{array}{l}
{{{a^2}} \mathord{\left/
 {\vphantom {{{a^2}} {{{\sin }^2}{\varphi_{\rm{c}}}}}} \right.
 \kern-\nulldelimiterspace} {{{\sin }^2}{\varphi_{\rm{c}}}}},\;\;\;{\rm{if}}\;\psi _{i,j}^{{\rm{LoS}},{\rm{in}}} \le {\varphi _{\rm{c}}}, \\
 [1mm]
0,\;\;\;\;\;\;\;\;\;\;\;\;\;\;\;\;{\rm{if}}\;\psi _{i,j}^{{\rm{LoS}},{\rm{in}}} > {\varphi _{\rm{c}}},
\end{array} \right.
\end{equation}
where $a$ denotes the refractive index, and $\varphi_{\rm{c}}$ is half of FoV at a receiver.

For a first refection link \cite{2004-NLOS},
\begin{equation}
\label{NLOS-gain}
\begin{aligned} 
dh_{i,j}^{{\rm{NLoS}}} = &{{\eta (L_a + 1)\delta \varpi {T}\rho } \over {2\pi d\,_{1,i,j}^2d\,_{2,i,j}^2}}J{\cos ^{L_a}}(\psi _{i,j}^{{\rm{NLoS,ir}}}) 
g(\psi _{i,j}^{{\rm{NLoS,in}}})\vspace{1ex}\cos (\psi _{i,j}^{{\rm{NLoS,in}}})d{A_{walls}},
\end{aligned} 
\end{equation}
where $J=\cos ({\omega _1})\cos ({\omega _2})$, $\omega _1$ and $\omega _2$ denote the angle of irradiance to a reflective point and the angle of irradiance to the receiver, respectively, $\rho$ is the reflectance factor, and $d{A_{walls}}$ denotes a small reflective area. Similar to the LoS link, $d_{1, i, j}$ is the distance between user $j$ and a reflective point, and $d_{2, i, j}$ is the distance between the reflective point and AP $i$. In addition, $\psi _{i,j}^{{\rm{NLoS,in}}}$ and $\psi _{i,j}^{{\rm{NLoS,ir}}}$ are the angle of incidence and the angle of irradiance of the first reflection link between optical AP $i$ and user $j$, respectively.

\section{Effective secrecy rate for the multi-user multi-AP VLC network}
\label{section-pri-rate}
In this section, we derive the ESR as a function of the secure cell formation parameters to characterize the delay-constrained secrecy rate. First, the artificial noise technique is adopted for each secure cell to enhance secrecy and the artificial noise aided achievable secrecy rate is derived. Further, considering the impact of the statistical delay requirement on the secrecy rate, we derive the ESR based on EC theory.

\subsection{Artificial noise aided achievable secrecy rate}
\label{subsection-secrecy rate}
Since the large number of NLoS wiretap links complicates the problem of finding an efficient beamformer, we adopt the artificial noise technique to enhance secrecy. 
In timeslot $t$, the downlink signal transmitted in the secure cell of user $j$ is divided into the useful signal $s^u_j(t)$ and independent jamming signals $s^a_{j_l}(t), \forall l \in \left\{ {1,2, \cdots ,\left| {{\Omega _j}} \right| - 1} \right\}$. 
To guarantee the jamming signals are transmitted in the nullspace of user $j$, we define ${{\bf{\hat \Gamma }}_j}(t) \in \mathbb{R}^{{\left| {{\Omega _j}} \right|} \times {(\left| {{\Omega _j}} \right|-1)}} $ as a matrix whose columns ${\hat \Gamma _{{j_1}}}(t),{\hat \Gamma _{{j_2}}}(t), \cdots ,{\hat \Gamma _{{j_{| {{\Omega _j}}| - 1}}}}(t)$ constitute a basis for the nullspace of ${\bf{h}}_j^T(t)$ and are normalized such that ${\left\| {{{\hat \Gamma }_{{j_l}}}}(t) \right\|_1} = 1,\;\forall l \in \left\{ {1,2, \cdots ,\left| {{\Omega _j}} \right| - 1} \right\}$.  
Then, the transmitted signal can be expressed by
\begin{equation}
\begin{aligned}
{{\bf{s}}_j}(t) = {{\bf{w}}_j}(t)s_j^u(t) + \sum\limits_{l = 1}^{\left| {{\Omega _j}} \right| - 1} {{{\hat \Gamma }_{{j_l}}}(t)s_{{j_l}}^a(t)},
\end{aligned}
\end{equation}
where ${\bf{w}}_j(t)$ is a ${\left| {{\Omega _j}} \right| \times {\rm{1}}}$ precoding vector satisfying ${{\left\| {{{\bf{w}}_j}(t)} \right\|}_\infty } \le 1.$
Considering the peak power constraint in (\ref{equ-peak power constraint}), we define $\alpha_j(t) \in [0,1]$ as the power separation factor of user $j$. The fraction of optical power $A$ allocated to the useful signal is $\alpha_j(t) A$, while the fraction of $A$ allocated to the jamming signals is $(1-\alpha_j(t))A$. To be fair to each eavesdropper, the $(1-\alpha_j(t))A$ power fraction is divided equally among the available $\left| {{\Omega _j}} \right|-1$ nullspace directions.
Therefore, the constraints on signals are expressed as 
\begin{equation}
{\left| {s_j^u(t)} \right| \le {\alpha _j}(t)A,\;\;\;\;\;\left| {s_{{j_l}}^a(t)} \right| \le \frac{{(1 - {\alpha _j}(t))A}}{{\left| {{\Omega _j}} \right| - 1}}.}
\end{equation} 
In Sec. \ref{section-Lya}, precoding vector ${{\bf{w}}_j}(t)$ and power separation factor $\alpha_j(t)$ are treated as optimization variables in our stochastic optimization problem.

In this paper, when we form secure cells, users who may interfere with each other will be scheduled in different timeslots. Therefore, in timeslot $t$, the received signal of user $j$ in its secure cell is interference-free, which is expressed as 
\begin{equation}
\label{recev-signal-AN}
{y_j^{\rm{ }}(t)} = {\bf{h}}_j^T(t){{\bf{w}}_j(t)}{s^u_j(t)} + {n_j(t)}.
\end{equation}
We assume all users in the same room are potential eavesdroppers of user $j$, no matter they are scheduled or not in timeslot $t$. For eavesdropper $k \in {\Psi _j} = \left\{ {k \in \Phi \left| {\;k \ne j} \right.} \right\}$, the received eavesdropped signal intended to user $j$ is expressed as 
\begin{equation}
\begin{aligned}
\label{recev-signal-AN-eve}
y_{j,k}^{\rm{e}}(t) = {\rm{ }}{\left( {{\bf{h}}_{j,k}^{\rm{e}}(t)} \right)^T}{{\bf{w}}_j}(t)s_j^u(t) + {\left( {{\bf{h}}_{j,k}^{\rm{e}}(t)} \right)^T}\sum\limits_{l = 1}^{\left| {{\Omega _j}} \right| - 1} {{{\hat \Gamma }_{{j_l}}}(t)s_{{j_l}}^a(t)}  + n_{j,k}^{\rm{e}}(t).
\end{aligned}
\end{equation}

Considering the channel blockage of user $k$, the elements of ${{\bf{h}}_{j,k}^{{\rm{e}}}}(t)$ can be expressed as
\begin{equation}
\label{case1-gain-prob-eve}
h_{i,j,k}^{{\rm{e}}}{\rm{ = }}\xi_k(t) h_{i,k}^{{\rm{LoS}}}+\int_{{\rm{walls}}} {1\;} dh_{i,k}^{{\rm{NLoS}}},\;i \in {\Omega _j},
\end{equation}
which means user $k$ can eavesdrop user $j$'s signal through the LoS and the NLoS links combined or through the NLoS links if the LoS link is blocked. 

Since there is no collusion among eavesdroppers, the achievable secrecy rate of user $j$ in the multi-user multi-AP VLC network is defined by the strongest eavesdropper\cite{Oxford-2018-WCL}, as described in the following theorem. 
\begin{theorem}
\label{theorem1} 
For user $j$ in an indoor multi-AP multi-user VLC network, when all the other users in the same room are potential eavesdroppers of user $j$, the artificial noise aided achievable secrecy rate of user $j$ is expressed as 
\begin{equation}
\label{pri-capacity-nonAN}
{\tilde R}_j^{{\rm{s}}}(\alpha_j(t),{{\bf{w}}_j}(t),{\bf{\Pi }}_j(t)) \buildrel \Delta \over = \mathop {\min }\limits_{k \in {\Psi _j}}  \; {{\mathds{1}}_{\left({{{\left\| {{{\bf{\Pi }}_j}(t)} \right\|}_\infty } \ne 0} \right)}}R_{j,k}^{\rm{s}}(\alpha_j(t),{{\bf{w}}_j}(t)),
\end{equation}
where ${\mathds{1}}_{(\cdot)}$ is a binary indicator function, $R_{j,k}^{{\rm{s}}}(\alpha_j(t),{{\bf{w}}_j}(t))$ is the artificial noise aided achievable secrecy rate of user $j$ against eavesdropper $k$. For the wiretap model (\ref{recev-signal-AN-eve}), a lower bound on $R_{j,k}^{{\rm{s}}}(\alpha_j(t),{{\bf{w}}_j}(t))$ can be derived as 
\begin{equation}
\label{result-appendixA}
\begin{array}{*{20}{l}}
{R_{j,k}^{\rm{s}}(\alpha_j(t),{{\bf{w}}_j}(t) )}\\
{ \ge \frac{1}{2}\log (4{\bf{w}}_j^T(t){\bf{h}}_j^{}(t){\bf{h}}_j^T(t){{\bf{w}}_j}(t)\alpha_j^2(t)A^2 + 2\pi e\sigma^2) - \frac{1}{2}\log (2\pi e\sigma^2)}\vspace{1ex} \\
{ - \frac{1}{2}\log \left[ {\frac{2\pi e}{3}({\bf{w}}_j^T(t){\bf{h}}_{j,k}^{{\rm{e}}}(t){{\left( {{\bf{h}}_{j,k}^{{\rm{e}}}} (t)\right)}^T}{{\bf{w}}_j(t)}\alpha_j^2(t) A^2} 
 { + \sum\limits_{l = 1}^{\left| {{\Omega _j}} \right| - 1} {\hat \Gamma _{{j_l}}^T{\bf{h}}_{j,k}^{\rm{e}}{{\left( {{\bf{h}}_{j,k}^{\rm{e}}} \right)}^T}{{\hat \Gamma }_{{j_l}}}\frac{{{{(1 - {\alpha _j}(t))}^2}{A^2}}}{{{{\left( {\left| {{\Omega _j}} \right| - 1} \right)}^2}}}}  + 3{\sigma ^2})} \right]}\vspace{1ex} \\
{ + \frac{1}{2}\log \left[ {4\sum\limits_{l = 1}^{\left| {{\Omega _j}} \right| - 1} {\hat \Gamma _{{j_l}}^T{\bf{h}}_{j,k}^{\rm{e}}{{\left( {{\bf{h}}_{j,k}^{\rm{e}}} \right)}^T}{{\hat \Gamma }_{{j_l}}}\frac{{{{(1 - {\alpha _j}(t))}^2}{A^2}}}{{{{\left( {\left| {{\Omega _j}} \right| - 1} \right)}^2}}}}  + 2\pi e{\sigma ^2}} \right],}
\end{array}
\end{equation}
where $\log (\cdot)$ denotes the natural logarithm. 

\begin{IEEEproof}
See Appendix \ref{Appendix-A}.
\end{IEEEproof}
\end{theorem}



\begin{figure}[!t]
\centering
\resizebox{3.5in}{!}{%
\begin{tikzpicture}

\begin{axis}[%
width=4.241in,
height=2.993in,
at={(0.711in,0.443in)},
scale only axis,
xmin=0,
xmax=1,
xlabel style={font=\color{white!15!black}},
xlabel={$\text{Power separation factor }\alpha(t)$},
ymin=0,
ymax=35000000,
ylabel style={font=\color{white!15!black}, align=center},
ylabel={Average achievable secrecy transmission rate per user (bits/s)},
axis background/.style={fill=white},
legend style={at={(axis cs:0.02,500000)}, anchor=south west, legend cell align=left, align=left, draw=white!15!black}
]
\addplot [color=blue]
  table[row sep=crcr]{%
0	0\\
0.01	7654645.52646811\\
0.02	12377829.6814718\\
0.03	15247906.6050569\\
0.04	17303134.8021197\\
0.05	18902859.9207167\\
0.06	20210038.9433684\\
0.07	21314488.6557806\\
0.08	22270357.8520567\\
0.09	23112308.3866285\\
0.1	23864059.217162\\
0.11	24524636.0031558\\
0.12	25105727.1414304\\
0.13	25634486.9197296\\
0.14	26118603.842864\\
0.15	26564200.0940127\\
0.16	26976227.7164567\\
0.17	27358745.9818889\\
0.18	27715120.653121\\
0.19	28048170.3886429\\
0.2	28360276.4415417\\
0.21	28653466.2769112\\
0.22	28929478.2722228\\
0.23	29189812.4425845\\
0.24	29435770.6705188\\
0.25	29668488.936516\\
0.26	29888963.3718565\\
0.27	30098071.4835265\\
0.28	30296589.5656721\\
0.29	30485207.0697932\\
0.3	30664538.5283715\\
0.31	30835133.4948276\\
0.32	30997484.863652\\
0.33	31152035.8592699\\
0.34	31299185.9243972\\
0.35	31439295.6938248\\
0.36	31572691.2044996\\
0.37	31699667.4651041\\
0.38	31820491.4863179\\
0.39	31935404.8552976\\
0.4	32044625.9236564\\
0.41	32148351.6666321\\
0.42	32241776.1903607\\
0.43	32329435.0791662\\
0.44	32412184.842308\\
0.45	32490183.5713561\\
0.46	32563575.9314426\\
0.47	32632493.9364085\\
0.48	32697057.6164273\\
0.49	32757375.5932893\\
0.5	32813545.576313\\
0.51	32865654.7899803\\
0.52	32913780.3428126\\
0.53	32957989.5456745\\
0.54	32998340.1865519\\
0.55	33034880.7678811\\
0.56	33067650.7116523\\
0.57	33096680.5367568\\
0.58	33121992.0123468\\
0.59	33143598.2903103\\
0.6	33161504.0192864\\
0.61	33175705.4419429\\
0.62	33186190.4764522\\
0.63	33192938.7822203\\
0.64	33195921.8088838\\
0.65	33195102.8263592\\
0.66	33190436.9322476\\
0.67	33181871.0310961\\
0.68	33156755.1715224\\
0.69	33124319.6689661\\
0.7	33085659.3519368\\
0.71	33040529.7560329\\
0.72	32988666.5777206\\
0.73	32929784.0259017\\
0.74	32863572.7410126\\
0.75	32789697.1541306\\
0.76	32707792.1194324\\
0.77	32617458.6006675\\
0.78	32518258.1203906\\
0.79	32409705.5811736\\
0.8	32291259.9283499\\
0.81	32162311.9251048\\
0.82	32022168.0238363\\
0.83	31803935.9638193\\
0.84	31555727.4388386\\
0.85	31287458.0422762\\
0.86	30997034.7820155\\
0.87	30681888.5665305\\
0.88	30338811.3903969\\
0.89	29963720.6365638\\
0.9	29551308.9378127\\
0.91	29094507.7531334\\
0.92	28583634.2366311\\
0.93	28004970.2780119\\
0.94	27338254.1213021\\
0.95	26551908.3772175\\
0.96	25593014.6596887\\
0.97	24363146.65298\\
0.98	22646940.6555668\\
0.99	19289004.8710912\\
1	11535239.8511367\\
};
\addlegendentry{with artificial noise and ${\bf{w}}(t)$ are all-ones vectors}

\addplot [color=green, dashdotted]
  table[row sep=crcr]{%
0	0\\
0.01	4358790.15892179\\
0.02	8682101.65918103\\
0.03	11465764.2908138\\
0.04	13490073.8236703\\
0.05	15075808.8925625\\
0.06	16377813.9474579\\
0.07	17481690.9606333\\
0.08	18439528.0582668\\
0.09	19285314.6412149\\
0.1	20042433.8962164\\
0.11	20727650.4500988\\
0.12	21352797.065441\\
0.13	21927113.8854297\\
0.14	22458607.1426593\\
0.15	22953116.9458686\\
0.16	23415354.148943\\
0.17	23849172.6285226\\
0.18	24257764.1503154\\
0.19	24643800.7922833\\
0.2	24996838.2196232\\
0.21	25324007.9898324\\
0.22	25634035.1110924\\
0.23	25928406.7003124\\
0.24	26208421.7130252\\
0.25	26475220.8994659\\
0.26	26729810.9165985\\
0.27	26973083.9211801\\
0.28	27205833.6310159\\
0.29	27428768.5984103\\
0.3	27642523.2626263\\
0.31	27847667.217511\\
0.32	28044713.0330561\\
0.33	28234122.8963237\\
0.34	28416314.2814169\\
0.35	28591664.8154036\\
0.36	28760516.4740119\\
0.37	28923179.2150992\\
0.38	29079934.13759\\
0.39	29231036.237485\\
0.4	29376716.8196818\\
0.41	29517185.6139898\\
0.42	29652632.635313\\
0.43	29783229.8210881\\
0.44	29909132.4733777\\
0.45	30030480.5282735\\
0.46	30147399.6712737\\
0.47	30260002.313894\\
0.48	30368388.4438487\\
0.49	30472646.3585769\\
0.5	30572853.2896378\\
0.51	30669075.9234629\\
0.52	30761370.8221092\\
0.53	30849784.7459331\\
0.54	30934354.8784807\\
0.55	31015108.9523173\\
0.56	31092065.2729688\\
0.57	31165232.6365871\\
0.58	31234610.1353438\\
0.59	31300186.8428729\\
0.6	31361941.3702753\\
0.61	31419841.2812429\\
0.62	31473842.3526929\\
0.63	31523887.6648915\\
0.64	31569906.5023282\\
0.65	31611813.0435136\\
0.66	31649504.8143496\\
0.67	31682860.8756741\\
0.68	31711739.710912\\
0.69	31735976.7743438\\
0.7	31755381.6541666\\
0.71	31769734.7970241\\
0.72	31778783.7316974\\
0.73	31782238.7186405\\
0.74	31779767.7382379\\
0.75	31770990.7127898\\
0.76	31755472.8333317\\
0.77	31732716.8293377\\
0.78	31702153.9721778\\
0.79	31663133.5340599\\
0.8	31614910.3207014\\
0.81	31556629.7386141\\
0.82	31487309.6158801\\
0.83	31405817.6199211\\
0.84	31310842.5293639\\
0.85	31200856.6941182\\
0.86	31074065.5509985\\
0.87	30928337.6996432\\
0.88	30708920.3960686\\
0.89	30407676.1613292\\
0.9	30064279.9970893\\
0.91	29671139.759311\\
0.92	29218091.2593171\\
0.93	28690994.398325\\
0.94	28069240.9560151\\
0.95	27320996.580384\\
0.96	26393185.9967395\\
0.97	25187332.2579366\\
0.98	23488128.4804822\\
0.99	20668518.0862666\\
1	13534538.0958806\\
};
\addlegendentry{with artificial noise and ${\bf{w}}(t)$ are determined by MRT}

\addplot [color=red, dashed]
  table[row sep=crcr]{%
0	8919578.77263054\\
0.01	8919578.77263054\\
0.02	8919578.77263054\\
0.03	8919578.77263054\\
0.04	8919578.77263054\\
0.05	8919578.77263054\\
0.06	8919578.77263054\\
0.07	8919578.77263054\\
0.08	8919578.77263054\\
0.09	8919578.77263054\\
0.1	8919578.77263054\\
0.11	8919578.77263054\\
0.12	8919578.77263054\\
0.13	8919578.77263054\\
0.14	8919578.77263054\\
0.15	8919578.77263054\\
0.16	8919578.77263054\\
0.17	8919578.77263054\\
0.18	8919578.77263054\\
0.19	8919578.77263054\\
0.2	8919578.77263054\\
0.21	8919578.77263054\\
0.22	8919578.77263054\\
0.23	8919578.77263054\\
0.24	8919578.77263054\\
0.25	8919578.77263054\\
0.26	8919578.77263054\\
0.27	8919578.77263054\\
0.28	8919578.77263054\\
0.29	8919578.77263054\\
0.3	8919578.77263054\\
0.31	8919578.77263054\\
0.32	8919578.77263054\\
0.33	8919578.77263054\\
0.34	8919578.77263054\\
0.35	8919578.77263054\\
0.36	8919578.77263054\\
0.37	8919578.77263054\\
0.38	8919578.77263054\\
0.39	8919578.77263054\\
0.4	8919578.77263054\\
0.41	8919578.77263054\\
0.42	8919578.77263054\\
0.43	8919578.77263054\\
0.44	8919578.77263054\\
0.45	8919578.77263054\\
0.46	8919578.77263054\\
0.47	8919578.77263054\\
0.48	8919578.77263054\\
0.49	8919578.77263054\\
0.5	8919578.77263054\\
0.51	8919578.77263054\\
0.52	8919578.77263054\\
0.53	8919578.77263054\\
0.54	8919578.77263054\\
0.55	8919578.77263054\\
0.56	8919578.77263054\\
0.57	8919578.77263054\\
0.58	8919578.77263054\\
0.59	8919578.77263054\\
0.6	8919578.77263054\\
0.61	8919578.77263054\\
0.62	8919578.77263054\\
0.63	8919578.77263054\\
0.64	8919578.77263054\\
0.65	8919578.77263054\\
0.66	8919578.77263054\\
0.67	8919578.77263054\\
0.68	8919578.77263054\\
0.69	8919578.77263054\\
0.7	8919578.77263054\\
0.71	8919578.77263054\\
0.72	8919578.77263054\\
0.73	8919578.77263054\\
0.74	8919578.77263054\\
0.75	8919578.77263054\\
0.76	8919578.77263054\\
0.77	8919578.77263054\\
0.78	8919578.77263054\\
0.79	8919578.77263054\\
0.8	8919578.77263054\\
0.81	8919578.77263054\\
0.82	8919578.77263054\\
0.83	8919578.77263054\\
0.84	8919578.77263054\\
0.85	8919578.77263054\\
0.86	8919578.77263054\\
0.87	8919578.77263054\\
0.88	8919578.77263054\\
0.89	8919578.77263054\\
0.9	8919578.77263054\\
0.91	8919578.77263054\\
0.92	8919578.77263054\\
0.93	8919578.77263054\\
0.94	8919578.77263054\\
0.95	8919578.77263054\\
0.96	8919578.77263054\\
0.97	8919578.77263054\\
0.98	8919578.77263054\\
0.99	8919578.77263054\\
1	8919578.77263054\\
};
\addlegendentry{without artificial noise}

\end{axis}

\end{tikzpicture}%
}
\vspace*{-7mm}
\caption{Achievable secrecy rate with and without artificial noise, where 10 users randomly distributed in a  $16\rm{m}\times16\rm{m}\times2.5\rm{m}$ room and the available bandwidth is 20MHz.}
\vspace*{-7mm}
\label{Fig-pri-rate-with-out-AN}
\end{figure}
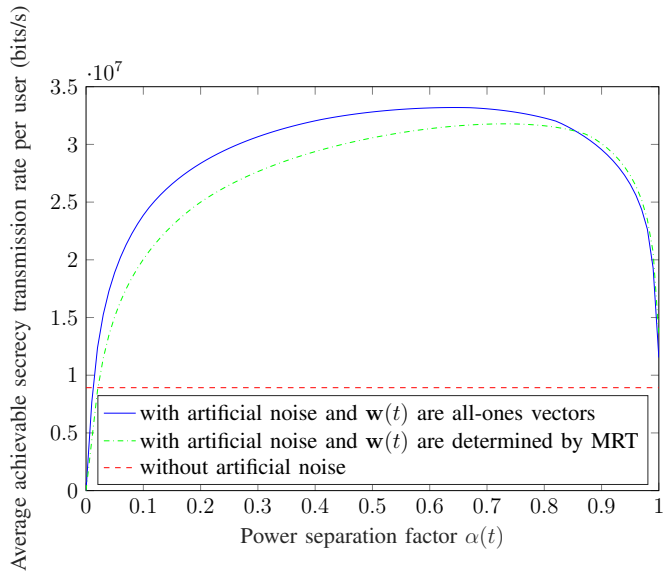

Based on Theorem \ref{theorem1}, Fig. \ref{Fig-pri-rate-with-out-AN} shows a comparison between achievable secrecy rates with and without artificial noise under the assumption that all users use the same $\alpha(t)$. 
Users' precoding vectors are determined by maximum ratio transmission (MRT) or all-ones vectors. 
Due to the constraint ${{\left\| {{{\bf{w}}}(t)} \right\|}_\infty } \le 1$, all-ones precoding vectors have the maximum norm, which means that ${\bf{w}}(t)$ is transmitted with maximum power.
Fig. \ref{Fig-pri-rate-with-out-AN} verifies that the precoding and jamming can enhance the secrecy rates of users effectively. 
Particularly, when the jamming power is dominating (i.e. $\alpha(t)$ is small), using the all-ones vector leads to improve the achievable rate at the legitimate user, while the effect of that on the eavesdropper's rate is small since it is dominated by the jamming signal. Therefore, all-ones vectors perform better than MRT vectors for small $\alpha(t)$ in Fig. \ref{Fig-pri-rate-with-out-AN}.
When the jamming power is relatively small (i.e. $\alpha(t)$ is large), increasing the norm of ${\bf{w}}(t)$ would improve the rate of both the legitimate user and the eavesdropper with approximately the same level.
In this case, using the MRT vector (i.e. control the direction of the vector ${\bf{w}}(t)$), the legitimate user's rate would be improved more than the rate at the eavesdropper. 
From the above, the power separation factor $\alpha(t)$ and the precoding vector ${\bf{w}}(t)$ should be jointly optimized, which is considered in Sec. \ref{section-Lya}.

\vspace*{-5mm}
\subsection{Effective secrecy rate}
\label{section-EC}
In this paper, EC is introduced to mathematically characterize the impact of heterogeneous statistical delay QoS requirements on secrecy rates. Based on the queuing model, the buffer of user $j$ can be treated as a queue with stochastic arrivals. The service process of the queue is the secrecy transmission provided by VLC system. The QoS exponent $\theta _j$ of user $j$ is used to depict the delay bound violation probability of a stable queuing system in the EC theory. From a queuing perspective, EC physically means the maximum constant packet arrival rate that can be supported by the service process of secure VLC transmission under the statistical delay limit characterized by $\theta _j$. Correspondingly, there is another concept named effective bandwidth (EB), which represents the minimum constant service rate that the arrival process needs subject to the statistical delay constraint $\theta _j$. According to the EC and EB theory, the statistical delay QoS of user $j$ is guaranteed if and only if the EC of user $j$ is not smaller than EB.

Since user-centric secure cell formation is adjusted dynamically over timeslots and the LoS links are blocked randomly, the change of the achievable secrecy rate of user $j$ (i.e. the service process of the buffer of user $j$ from the queueing perspective) is a stochastic process. Then, we derive the ESR of the service process of user $j$ as \cite{Wu-2003}
\begin{equation}
\label{EC-expectation}
R_j^{{\rm{es}}}\left( {\alpha_j(t),{{\bf{w}}_j}(t), {\bf{\Pi }}_j(t)} \right) =  - \frac{1}{{{\theta _j}}}\log \mathbb{E} \left[ {{e^{ - {\theta _j}  {\tilde R}_j^{\rm{s}}(\alpha_j(t),{{\bf{w}}_j}(t),{\bf{\Pi }}_j(t))}}} \right],
\end{equation}
where $\mathbb{E} \left[ \cdot \right]$ denotes the expectation.

Characterizing the statistical delay guaranteed secrecy rate, ESR is a long-term metric, related to multiple observations of queuing behaviour on a large time scale.
However, to guarantee statistical delay QoS, we need to adjust the service process by operating the system on a small time scale, such as dynamically allocating the system resources, scheduling over timeslots, etc. 
To build a bridge between the large time granularity QoS requirement and the small-time granularity system operation, we approximately express the ESR as
\begin{equation}
\label{EC-limit}
R_j^{{\rm{es}}}\left( {{\alpha _j}(t),{{\bf{w}}_j}(t),{{\bf{\Pi }}_j}(t)} \right) \approx  - \frac{1}{{{\theta _j}}}\log \left[ {\frac{1}{t}\sum\limits_{\tau  = 0}^{t - 1} {{e^{ - {\theta _j}\tilde R_j^{\rm{s}}({\alpha _j}(\tau ),{{\bf{w}}_j}(\tau ),{{\bf{\Pi }}_j}(\tau ))}}} } \right],
\end{equation}
where the time average of ${{e^{ - {\theta _j}  {\tilde R}_j^{\rm{s}}(\alpha_j(t),{{\bf{w}}_j}(t),{\bf{\Pi }}_j(t))}}}$ is used to estimate its expectation.

\begin{figure}[!t]
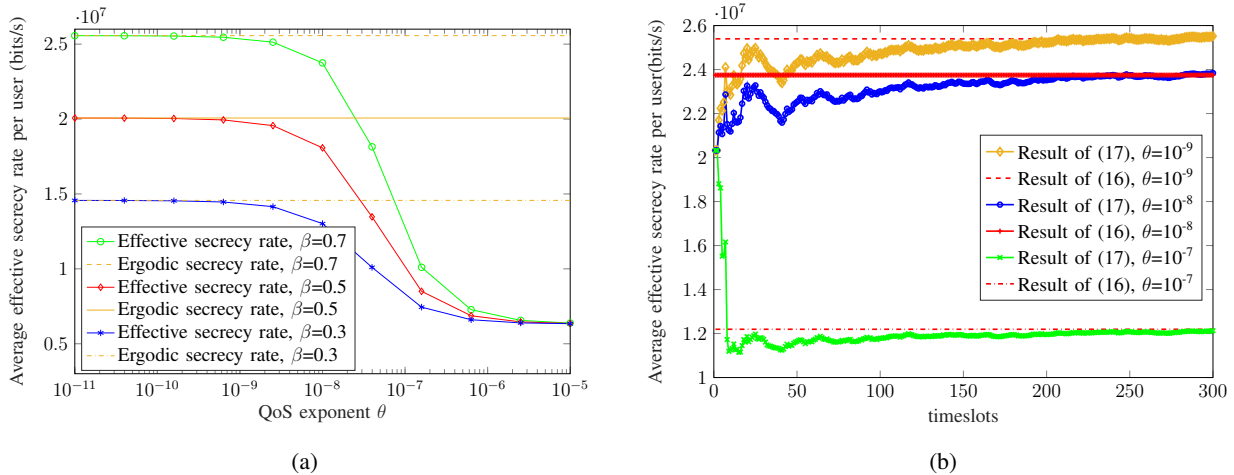

\centering
\subfloat[\label{Fig-EC-QoS}]{%
\resizebox{3.1in}{!}{%
%
}
}
\vspace*{-2mm}
\caption{(a) The average effective secrecy rate per user for various QoS exponents in different blockage conditions. (b) The effective secrecy rates calculated by (\ref{EC-expectation}) and (\ref{EC-limit}) for various QoS exponents.}
\vspace*{-5mm}
\end{figure}

Fig. \ref{Fig-EC-QoS} shows the ESRs and ergodic secrecy rates versus the QoS exponent under different blockage probabilities, where the randomness of the service process is caused by blockages. 
The figure calculates the ESR and the ergodic secrecy rate according to (\ref{EC-expectation}) and (\ref{pri-capacity-nonAN}), respectively. 
From Fig. \ref{Fig-EC-QoS}, we can see that when the delay constraint is quite loose (i.e. QoS exponent $\theta \to 0$), the ESR converges to the ergodic secrecy rate. 
As the statistical delay requirement increases (i.e. QoS exponent becoming larger), the required transmission rates compromise the secrecy performance, which leads to monotonically decreasing the ESR.
In other words, a user with a stringent delay requirement has to settle for a lower delay-guaranteed secrecy rate, compared to one with a loose delay constraint. 
The figure also shows that a high ESR is obtained when the channel is in good condition, which is indicated by a low blockage probability.  
In addition, Fig. \ref{Fig-EC-QoS-con} verifies the equivalence of (\ref{EC-expectation}) and (\ref{EC-limit}). 
The randomness of the service process is caused by blockage. 
Fig. \ref{Fig-EC-QoS-con} shows that for different QoS exponents, the effective secrecy rate calculated by (\ref{EC-limit}) can approximately converge to the result calculated by (\ref{EC-expectation}).

\definecolor{mycolor1}{rgb}{0.92941,0.69412,0.12549}%

\section{Problem formulation and the solution}
\label{section-Lya}
\subsection{The long-term VLC cell formation optimization problem}
In the previous sections, the achievable secrecy rate of each user was derived and the heterogeneous statistical delay QoS requirement was embedded in the ESR of each user. To form user-centric secure cells with various delay QoS guarantees, we adopt a user-centric principle and aim to optimize user scheduling, the intensity of artificial noise and the precoding vector of each secure cell. The cell formulation problem is formulated as a sum ESR maximization problem, which is expressed as follows
\begin{subequations}
\label{op-longterm}
\begin{alignat}{2}
&\!\max_{\bm{\alpha}(t),{{\bf{w}}}(t),{{\bf{\Pi }}(t)}} &\qquad& \sum\limits_{j = 1}^{{K_u}} R_j^{{\rm{es}}}\left( {\alpha_j(t),{{\bf{w}}_j}(t),{\bf{\Pi }}_j(t)} \right) \label{op-longterm-objective}\\
&\text{subject to}\;(\text{s.t.}) &      & R_{j }^{{\rm{es}}}(\alpha_j(t),{{\bf{w}}_j}(t),{\bf{\Pi }}_j(t)) \ge B^{\rm{e}}_{j },\;\;j = 1,...,{K_u},\label{op-longterm-st-EC} \\
&                  &      & {\pi _{i,j}}(t) \in \{ 0,1\},\label{op-longterm-st-pai} \\
&                  &      & \sum\limits_{i = 1}^{{K_a}} {{\pi _{i,j}}(t)} \sum\limits_{i = 1}^{{K_a}} {{\pi _{i,k}}(t)}  = 0,\;\;j,k \in \left\{ {1, \ldots ,{K_u}|{\Omega _j} \cap {\Omega _k} \ne \emptyset } \right\},\label{op-longterm-st-MISO} \\
&                  &      & 0 \le \alpha_j(t)  \le 1, j = 1,...,{K_u},\label{op-longterm-st-alpha} \\
&                  &      & {\left\| {{{\bf{w}}_j}(t)} \right\|_\infty } \le 1, j = 1,...,{K_u}.\label{op-longterm-st-w}
\end{alignat}
\end{subequations}

In the optimization problem (\ref{op-longterm}), one of the optimization variables is the cell formation matrix ${\bf{\Pi }}(t)$. Indicated by constraint (\ref{op-longterm-st-pai}), ${\bf{\Pi }}(t)$ is a binary matrix for each timeslot. 
In this paper, ${\pi_{i,j}}(t)$ depends on the user scheduling in timeslot $t$ and the set of APs who are capable to serve user $j$, which is characterized by ${\Omega _j}$ in Sec.~\ref{section-system-description}. 
Constraint (\ref{op-longterm-st-MISO}) indicates that users who may interfere each other (i.e. ${\Omega _j} \cap {\Omega _k} \ne \emptyset$) cannot be scheduled in the same timeslot.
Optimization variables $\bm{\alpha}(t)$ and ${{\bf{w}}}(t)$ are secure cell parameters, which represent power separation factor and precoding vector of each user's secure cell, respectively.
The long-term objective function (\ref{op-longterm-objective}) is a system-level metric characterized by the sum ESR. 
Since the ESR represents the transmission rate with both statistical delay and security QoS guaranteed, the long-term constraint (\ref{op-longterm-st-EC}) is used to characterize the delay and secure transmission requirement of user $j$, where $B^{\rm{e}}_{j}$ is the effective bandwidth of user $j$.
According to the formula (\ref{EC-limit}), optimization problem (\ref{op-longterm}) is rewritten as
\begin{subequations}
\label{op-longterm2}
\begin{alignat}{2}
&\!\max_{\bm{\alpha}(t),{{\bf{w}}}(t),{{\bf{\Pi }}(t)}} &\qquad& \sum\limits_{j = 1}^{{K_u}} { - \frac{1}{{{\theta _j}}}\log \left[ {\mathop {\lim }\limits_{t \to \infty } \frac{1}{t}\sum\limits_{\tau  = 0}^{t - 1} {{e^{ - {\theta _j} {\tilde R}_j^{\rm{s}}(\alpha_j(\tau),{{\bf{w}}_j}(\tau),{\bf{\Pi }}_j(\tau))}}} } \right]} \label{op-longterm2-st-objective}\\
& \quad \quad \text{s.t.} &      & \mathop {\lim }\limits_{t \to \infty } \frac{1}{t}\sum\limits_{\tau  = 0}^{t - 1} {{e^{ - {\theta _j} {\tilde R}_j^{\rm{s}}(\alpha_j(\tau),{{\bf{w}}_j}(\tau),{\bf{\Pi }}_j(\tau))}}}  \le {e^{ - B^{\rm{e}}_{j} {\theta _j}}},\;\;j = 1,...,{K_u},\label{op-longterm2-st-EC} \\
&                  &      & {\pi _{i,j}}(t) \in \{ 0,1\},\label{op-longterm2-st-pai} \\
&                  &      & \sum\limits_{i = 1}^{{K_a}} {{\pi _{i,j}}(t)} \sum\limits_{i = 1}^{{K_a}} {{\pi _{i,k}}(t)}  = 0,\;\;j,k \in \left\{ {1, \ldots ,{K_u}|{\Omega _j} \cap {\Omega _k} \ne \emptyset } \right\}\label{op-longterm2-st-MISO} \\
&                  &      & 0 \le \alpha_j(t)  \le 1, j = 1,...,{K_u},\label{op-longterm2-st-alpha} \\
&                  &      & {\left\| {{{\bf{w}}_j}(t)} \right\|_\infty } \le 1, j = 1,...,{K_u}.\label{op-longterm2-st-w}
\end{alignat}
\end{subequations}

Optimization problem (\ref{op-longterm2}) aims to optimize the secure cell formation in each timeslot to achieve the long-term objective function and satisfy the long-term constraints. To solve an long-term optimization problem, the decision of ${\bm{\alpha}}(t)$, ${\bf{w}}(t)$ and ${\bf{\Pi}}(t)$ in timeslot $t$ needs to be made before observing the entire values of objective function and constraints. Mathematically, uncertainty is brought into the optimization problem by the long-term functions, and problem (\ref{op-longterm2}) is a stochastic optimization problem, which cannot be solved by conventional methodological approaches for deterministic optimization problems. To decompose the long-term stochastic optimization problem (\ref{op-longterm2}) into several short-term deterministic optimization problems, we employ the Lyapunov stochastic optimization framework in this paper. With the help of Lyapunov virtual queues, we formulate the per-slot optimization problems based on DPP method. By working out the per-slot optimization problems, we can obtain an adaptive user-centric secure cell formation method, in which APs allocation and intensities of artificial noise are adjusted dynamically over timeslots according to the priori values of objective function and constraints.

\subsection{Equivalent Lyapunov virtual queue stability problem}
An optimization problem with long-term constraints needs to be transformed into a queue stability problem \cite{Neely-book} according to the Lyapunov optimization approach. First, we construct the virtual discrete-time queue for each long-term constraint to measure how well the constraint is satisfied in the previous timeslots. To define the virtual queue of long-term constraint (\ref{op-longterm2-st-EC}), we rewrite (\ref{op-longterm2-st-EC}) as
\begin{equation}
\label{equ-equlivaent-ECEB}
\mathop {\lim }\limits_{t \to \infty } \frac{1}{t}\sum\limits_{\tau  = 0}^{t - 1} {{e^{ - {\theta _j} {\tilde R}_j^{\rm{s}}(\alpha_j(\tau),{{\bf{w}}_j}(\tau),{\bf{\Pi }}_j(\tau))}}}  - \mathop {\lim }\limits_{t \to \infty } \frac{1}{t}\sum\limits_{\tau  = 0}^{t - 1} {{e^{ - B^{\rm{e}}_{j} {\theta _j}}}}  \le 0.
\end{equation}

Then, we define a Lyapunov virtual queue for constraint (\ref{op-longterm2-st-EC}) as
\begin{equation}
\label{equ-virtual queue}
{F_j}(t + 1) = \max \left\{ {{F_j}(t) + {e^{ - {\theta _j} {\tilde R}_j^{\rm{s}}(\alpha_j(t),{{\bf{w}}_j}(t),{\bf{\Pi }}_j(t))}} - {e^{ - B^{\rm{e}}_{j}  {\theta _j}}},0} \right\}.
\end{equation}
The dynamic equation (\ref{equ-virtual queue}) describes the updating process of a virtual discrete-time queue, in which ${F_j}(t)$ is viewed as the virtual queue length with arrival rate ${e^{ - {\theta _j} {\tilde R}_j^{\rm{s}}(\alpha_j(t),{{\bf{w}}_j}(t),{\bf{\Pi }}_j(t))}}$ and service rate ${e^{ - B^{\rm{e}}_{j} {\theta _j}}}$. The queue length ${F_j}(t)$ is also called the backlog, which depends on the difference between arrival process and service process in the historical timeslots \cite{Neely-book}. In this paper, ${F_j}(t)$ implies the gap between the previous VLC transmission service process (until timeslot $t$) provided for user $j$ and the ideal transmission service process required by the long-term guarantee goal of user $j$. For the timeslot $t+1$, the backlog of the virtual queue of the last timeslot (i.e. ${F_j}(t)$) can be regarded as a kind of feedback, which reflects the history of secrecy and delay guarantee conditions. The continuous congestion of a virtual queue means the corresponding secrecy and delay guarantee are moving away from the long-term goal.

\begin{lemma}
\label{lemma3}
The long-term ESR constraint (\ref{op-longterm2-st-EC}) is satisfied if the corresponding virtual queue (\ref{equ-virtual queue}) is stabilized, i.e. $\mathop {\lim }\limits_{t \to \infty } {F_j}(t) < \infty $. In other words, the stability of the corresponding virtual queue implies the feasibility of the long-term ESR constraint.

\begin{IEEEproof}
Initialize ${F_j}(0)=0$ for all $j = 1,...,{K_u}$. From the virtual queue (\ref{equ-virtual queue}), we have
\begin{equation}
{F_j}(t + 1) \ge {F_j}(t) + {e^{ - {\theta _j} {\tilde R}_j^{\rm{s}}(\alpha_j(t),{{\bf{w}}_j}(t),{\bf{\Pi }}_j(t))}} - {e^{ - B^{\rm{e}}_{j} {\theta _j}}}.
\end{equation}
Hence,
\begin{equation}
{e^{ - {\theta _j} {\tilde R}_j^{\rm{s}}(\alpha_j(t),{{\bf{w}}_j}(t),{\bf{\Pi }}_j(t))}} - {e^{ - B^{\rm{e}}_{j} {\theta _j}}} \le {F_j}(t + 1)-{F_j}(t).
\end{equation}
Then, summing both the left-hand side and the right-hand side of the inequality from the timeslot 0 to the timeslot $t-1$, we get
\begin{equation}
\label{equ-proof}
\frac{1}{t}\sum\limits_{\tau {\rm{ = 0}}}^{t - 1} {\left( {{e^{ - {\theta _j} {\tilde R}_j^{\rm{s}}(\alpha_j(\tau),{{\bf{w}}_j}(\tau),{\bf{\Pi }}_j(\tau))}} - {e^{ - B^{\rm{e}}_{j} {\theta _j}}}} \right)}  \le \frac{{{F_j}(t)}}{t}.
\end{equation}
Then, dividing both sides of (\ref{equ-proof}) by $t$, and letting $t$ go to infinity, we get
\begin{equation}
\mathop {\lim }\limits_{t \to \infty } \sum\limits_{\tau {\rm{ = 0}}}^{t - 1} {\left( {{e^{ - {\theta _j} {\tilde R}_j^{\rm{s}}(\alpha_j(\tau),{{\bf{w}}_j}(\tau),{\bf{\Pi }}_j(\tau))}} - {e^{ - B^{\rm{e}}_{j} {\theta _j}}}} \right)}  \le \mathop {\lim }\limits_{t \to \infty } {{F_j}(t)}.
\end{equation}
Therefore, the inequality (\ref{equ-equlivaent-ECEB}) is satisfied when 
\begin{equation}
\mathop {\lim }\limits_{t \to \infty } \frac{{{F_j}(t)}}{t}=0
\end{equation}
for all $j= 1,...,{K_u}$. Equivalently, the virtual queue needs to satisfy $\mathop {\lim }\limits_{t \to \infty } {F_j}(t) < \infty $, which means the virtual queue is stabilized.
\end{IEEEproof}
\end{lemma}

According to Lemma \ref{lemma3}, the optimization problem (\ref{op-longterm2}) can be transformed into a virtual queue stability problem as follows
\begingroup
\allowdisplaybreaks
\begin{subequations}
\label{op-longterm3}
\begin{alignat}{2}
&\!\max_{\bm{\alpha}(t),{{\bf{w}}}(t),{{\bf{\Pi }}(t)}} &\qquad& \sum\limits_{j = 1}^{{K_u}} { - \frac{1}{{{\theta _j}}}\log \left[ {\mathop {\lim }\limits_{t \to \infty } \frac{1}{t}\sum\limits_{\tau  = 0}^{t - 1} {{e^{ - {\theta _j} {\tilde R}_j^{\rm{s}}(\alpha_j(\tau),{{\bf{w}}_j}(\tau),{\bf{\Pi }}_j(\tau))}}} } \right]} \label{op-longterm3-st-objective}\\
&  \quad \quad \text{s.t.} &      & \mathop {\lim }\limits_{t \to \infty } {F_j}(t) < \infty , j = 1,...,{K_u},\label{op-longterm3-st-EC} \\
&                  &      & {\pi _{i,j}}(t) \in \{ 0,1\},\label{op-longterm3-st-pai} \\
&                  &      & \sum\limits_{i = 1}^{{K_a}} {{\pi _{i,j}}(t)} \sum\limits_{i = 1}^{{K_a}} {{\pi _{i,k}}(t)}  = 0,\;\;j,k \in \left\{ {1, \ldots ,{K_u}|{\Omega _j} \cap {\Omega _k} \ne \emptyset } \right\},\label{op-longterm3-st-MISO} \\
&                  &      & 0 \le \alpha_j(t) \le 1, j = 1,...,{K_u},\label{op-longterm3-st-alpha} \\
&                  &      & {\left\| {{{\bf{w}}_j}(t)} \right\|_\infty } \le 1, j = 1,...,{K_u}.\label{op-longterm3-st-w}
\end{alignat}
\end{subequations}
\endgroup
\vspace*{-15mm}
\subsection{Per-slot drift-plus-penalty (DPP) minimization problem}
In this subsection, we adopt Lyapunov DPP method to tackle the queue stability problem (\ref{op-longterm3}) and form secure VLC cells. Since secure cells need to be formed with all users' QoS requirements taken into account, we need a comprehensive evaluation of users' delay guarantee conditions. Therefore, based on the Lyapunov virtual queue of each long-term constraint, we construct a quadratic Lyapunov function \cite{Neely-book} as a non-negative scalar measure of all virtual queues' congestion, which is given by
\begin{equation}
L(t) = \frac{1}{2}\sum\limits_{j = 1}^{{K_u}} {{F_j^2}{{(t)}}}.
\end{equation}
To keep the virtual queues stable by persistently pushing the Lyapunov function towards a lower congestion state, we derive the one-step Lyapunov drift between two successive timeslots as

\begin{equation}
\begin{aligned}
\Delta L(t)&= L(t + 1) - L(t)\\
&= \frac{1}{2}\sum\limits_{j = 1}^{{K_u}} {\left[ {{F_j^2}{{(t + 1)}} - {F_j^2}{{(t)}}} \right]} \\
&=\frac{1}{2}\sum\limits_{j = 1}^{{K_u}} {\left[ {\max {{\left\{ {{F_j}(t) + {e^{ - {\theta _j}\tilde R_j^{\rm{s}}(\alpha_j(t),{{\bf{w}}_j}(t),{\bf{\Pi }}_j(t))}} - {e^{ - {B^{\rm{e}}_j}{\theta _j}}},0} \right\}}^2} - {F_j}{{(t)}^2}} \right]} .\\
\end{aligned}
\end{equation}
An upper bound on $\Delta L(t)$ can be expressed as
\begin{equation}
\begin{aligned}
\Delta L(t) &\le \frac{1}{2}\sum\limits_{j = 1}^{{K_u}} {\left[ {{{\left[ {{e^{ - {\theta _j}\tilde R_j^{\rm{s}}(\alpha_j(t),{{\bf{w}}_j}(t),{\bf{\Pi }}_j(t))}} - {e^{ - {B^{\rm{e}}_j}{\theta _j}}}} \right]}^2} + 2{F_j}(t)\left[ {{e^{ - {\theta _j}\tilde R_j^{\rm{s}}(\alpha_j(t),{{\bf{w}}_j}(t),{\bf{\Pi }}_j(t))}} - {e^{ - {B^{\rm{e}}_j}{\theta _j}}}} \right]} \right]} \\
&= \frac{1}{2}\sum\limits_{j = 1}^{{K_u}} {{{\left[ {{e^{ - {\theta _j} {\tilde R}_j^{\rm{s}}(\alpha_j(t),{{\bf{w}}_j}(t),{\bf{\Pi }}_j(t))}} - {e^{ - B^{\rm{e}}_{j} {\theta _j}}}} \right]}^2}} + \sum\limits_{j = 1}^{{K_u}} {{F_j}(t)\left[ {{e^{ - {\theta _j} {\tilde R}_j^{\rm{s}}(\alpha_j(t),{{\bf{w}}_j}(t),{\bf{\Pi }}_j(t))}} - {e^{ - B^{\rm{e}}_{j} {\theta _j}}}} \right]} \\
&\le D + \sum\limits_{j = 1}^{{K_u}} {{F_j}(t)\left[ {{e^{ - {\theta _j} {\tilde R}_j^{\rm{s}}(\alpha_j(t),{{\bf{w}}_j}(t),{\bf{\Pi }}_j(t))}} - {e^{ - B^{\rm{e}}_{j} {\theta _j}}}} \right]}, 
\end{aligned}
\end{equation}
where $D > 0$ is a finite constant.

According to the Lyapunov optimization theory, the virtual queues can be stabilized by minimizing the Lyapunov drift \cite{Neely-book}. Therefore, optimizing the objective function (\ref{op-longterm3-st-objective}) with virtual queues constraints (\ref{op-longterm3-st-EC}) is equivalent to optimizing the Lyapunov DPP, in which the objective function (\ref{op-longterm3-st-objective}) is treated as a penalty. The Lyapunov DPP function can be written and bounded as
\begin{equation}
\begin{aligned}
&\Delta L(t) - \sum\limits_{j = 1}^{{K_u}} { - \frac{1}{{{\theta _j}}}\log \left[ {{e^{ - {\theta _j} {\tilde R}_j^{\rm{s}}(\alpha_j(t),{{\bf{w}}_j}(t),{\bf{\Pi }}_j(t))}}} \right]} \\
& \le D + \sum\limits_{j = 1}^{{K_u}} {{F_j}(t)\left[ {{e^{ - {\theta _j} {\tilde R}_j^{\rm{s}}(\alpha_j(t),{{\bf{w}}_j}(t),{\bf{\Pi }}_j(t))}} - {e^{ - B^{\rm{e}}_{j} {\theta _j}}}} \right]}  - \sum\limits_{j = 1}^{{K_u}} { - \frac{1}{{{\theta _j}}}\log \left[ {{e^{ - {\theta _j} {\tilde R}_j^{\rm{s}}(\alpha_j(t),{{\bf{w}}_j}(t),{\bf{\Pi }}_j(t))}}} \right]}.
\end{aligned}
\end{equation}

Therefore, the optimization problem (\ref{op-longterm3}) is transformed as
\begin{subequations}
\label{op-DPP}
\begin{alignat}{2}
&\!\min_{\bm{\alpha}(t),{\bf{w}}(t),{\bm{\Pi }(t)}} &\;\;&  \sum\limits_{j = 1}^{{K_u}} {\left\{ {\frac{1}{{{\theta _j}}}\log \left[ {{e^{ - {\theta _j}\tilde R_j^{\rm{s}}({\alpha _j}(t),{{\bf{w}}_j}(t),{{\bf{\Pi }}_j}(t))}}} \right] + {F_j}(t)\left[ {{e^{ - {\theta _j}\tilde R_j^{\rm{s}}({\alpha _j}(t),{{\bf{w}}_j}(t),{{\bf{\Pi }}_j}(t))}} - {e^{ - B_j^e{\theta _j}}}} \right]} \right\}}   \label{op-DPP-st-objective}\\
&  \quad \quad \text{s.t.} &      & {\pi _{i,j}}(t) \in \{ 0,1\},\label{op-DPP-st-pai} \\
&                  &      & \sum\limits_{i = 1}^{{K_a}} {{\pi _{i,j}}(t)} \sum\limits_{i = 1}^{{K_a}} {{\pi _{i,k}}(t)}  = 0,\;\;j,k \in \left\{ {1, \ldots ,{K_u}|{\Omega _j} \cap {\Omega _k} \ne \emptyset } \right\},\label{op-DPP-st-MISO} \\
&                  &      & 0 \le \alpha_j(t)  \le 1, j = 1,...,{K_u},\label{op-DPP-st-alpha} \\
&                  &      & {\left\| {{{\bf{w}}_j}(t)} \right\|_\infty } \le 1, j = 1,...,{K_u}.\label{op-DPP-st-w}
\end{alignat}
\end{subequations}

So far, we decompose the long-term stochastic problem (\ref{op-longterm2}) into several equivalent evolutionary per-slot DPP minimization problems as (\ref{op-DPP}). The optimization problem of timeslot $t+1$ is constructed adaptively according to the secrecy and delay guarantee conditions in priori timeslots. By solving problem (\ref{op-longterm2}) slot by slot, we can obtain the dynamic user-centric secure cell strategy. 
\subsection{The user-centric secure cell formation solution}
A mixed integer non-linear programming problem like (\ref{op-DPP}) is a kind of non-deterministic polynomial (NP) complete problems \cite{Chi-2018-COPP}, which has no exact solution in polynomial time. The joint optimization of user-centric scheduling and secure cell parameters imposes an extremely high computational complexity, which makes exhaustive search methods impractical. 
To solve the problem (\ref{op-DPP}) more efficiently, we decouple the problem (\ref{op-DPP}) into two sub-problems: the inner problem that is called intra-cell secure parameters optimization problem and the outer problem is called user-centric scheduling problem.
Specifically, with given ${{\bf{\Pi }}_j}$, the inner problem aims to determine the precoding vector ${\bf{w}}_j(t)$ and the power separation factor $\alpha_j(t)$ for each secure cell, while the outer problem is designed to find the optimal combination of simultaneous transmission secure cells in order to minimize the sum DPP by optimizing ${\bf{\Pi }}$.
For the sake of simplicity, the timeslot variable $t$ is omitted from our notations in this Subsection.
\subsubsection{The intra-cell secure parameters optimization problem}
Determined by user $j$'s scheduling, ${{\bf{\Pi }}_j}$ has only two possibilities: ${\pi_{i,j}}(t)=1$ for all $i \in {\Omega _j}$ if user $j$ is scheduled, and ${\pi_{i,j}}(t)=0 $ for all APs if user $j$ is not scheduled. 
To decompose the problem (\ref{op-DPP}), we first traverse all users and optimize the secure parameters ${\alpha}_j$ and ${\bf{w}}_j$ for the case that user $j$ is scheduled\footnote{When user $j$ is not scheduled, the achievable secrecy rate $\tilde R_j^{\rm{s}}({\alpha _j},{{\bf{w}}_j},{{\bf{\Pi }}_j}) = 0$, and the $j$th term of the objective function (\ref{op-DPP-st-objective}) identically equals to ${F_j}\left[ {1 - {e^{ - B_j^e{\theta _j}}}} \right]$, which means ${\alpha _j}$ and ${{\bf{w}}_j}$ are not required to be optimized in this case.}.
When corresponding ${{\bm{\Pi }}_j}$ is given, the secure cell of user $j$ can be formed like a downlink MISO system containing all APs in $\Omega_j$. 
Therefore, the intra-cell secure parameters optimization problem of user $j$ is expressed as\footnote{It is rational to optimize the secure parameters of user $j$'s cell independently because all other users are considered as eavesdroppers of user $j$ whether they are scheduled or not, in addition, the secure parameters of different cells are independent once constraint (\ref{op-DPP-st-MISO}) is satisfied. }

\vspace*{-2mm}
\begin{subequations}
\label{op-DPP-intracell}
\begin{alignat}{2}
&\!\min_{{\alpha}_j,{\bf{w}}_j} &\quad& { {\frac{1}{{{\theta _j}}}\log \left[ {{e^{ - {\theta _j}\tilde R_j^{\rm{s}}({\alpha _j},{{\bf{w}}_j},{{\bf{\Pi }}_j})}}} \right] + {F_j}\left[ {{e^{ - {\theta _j}\tilde R_j^{\rm{s}}({\alpha _j},{{\bf{w}}_j},{{\bf{\Pi }}_j})}} - {e^{ - B^{\rm{e}}_j{\theta _j}}}} \right]} }   \label{op-DPP-intracell-objective}\\
& \;\; \text{s.t.} &      & 0 \le \alpha_j \le 1, j = 1,...,{K_u},\\
&                  &      & {\left\| {{{\bf{w}}_j}} \right\|_\infty } \le 1, j = 1,...,{K_u}.
\end{alignat}
\end{subequations}

\subsubsection{The modified PSO algorithm}
We leverage PSO algorithm to solve problem (\ref{op-DPP-intracell}), which is a meta-heuristic algorithm which imitates swarms behaviour in birds flocking and fish schooling \cite{PSO1999}. In PSO, a swarm consisting of several particles is randomly generated during initialization. Each particle represents a potential solution of the optimization problem (\ref{op-DPP-intracell}). The position of each particle is a multidimensional variable including ${\alpha}_j$ and ${{\bf{w}}_j}$. The fitness value is defined to measure the optimality of particles' positions, which can be calculated by the objective function (\ref{op-DPP-intracell-objective}). Each particle searches for the minimum fitness value in the solution space by iteratively updating its position according to its own search experience and its companions' search experience. 
The updating equations of velocity and position are as follows
\begin{equation}
\label{equ-PSO-velocity}
\begin{aligned}
\bm{V}_{\lambda}^{n + 1} = &{\zeta^n}\bm{V}_{\lambda}^{n} + {c_1}{r_1}(\bm{P} _{\lambda,pbest}^{n} - \bm{P} _\lambda ^{n})+ {c_2}{r_2}(\bm{P}_{gbest}^{n} - \bm{P}_\lambda ^{n}), 
\end{aligned}
\end{equation}
\begin{equation}
\label{equ-PSO-position}
\bm{P}_{\lambda}^{n + 1} = \bm{P} _\lambda^{n}+\bm{V}_\lambda^{n+1}, 
\end{equation}
where $\bm{V}_{\lambda}^{n}$ and $\bm{P} _\lambda ^{n}$ are the velocity and position of particle $\lambda$ at the $n$th iteration, respectively. In addition, $\bm{P} _{\lambda,pbest}^{n}$ and $\bm{P}_{gbest}^{n}$ are the best position record of particle $\lambda$ and the entire swarm, respectively. 
Furthermore, $\zeta^n$ is the dynamic inertia weight, which is determined by $\zeta^n=0.9-0.5n/{n_{\rm{max}}}$, where ${n_{\rm{max}}}$ is the maximum iterations, $c_1=c_2=1$ are the learning factors, and $r_1$ and $r_2$ are random numbers uniformly distributed in the range $\left[ {0,1} \right]$.

When solving the non-convex problem (\ref{op-DPP-intracell}), we noticed that the conventional PSO algorithm converges prematurely since ${w_{i,j}=1}$ is always a component of an easily trapped local minimum solution. 
Therefore, we modify the PSO algorithm by considering the premature convergence of $\bm{P}_{gbest}^{n}$ in the iteration process. 
If the best position record of swarm $\bm{P}_{gbest}^{n}$ remains unchanged for $n_{\rm{th}}$ iterations, the algorithm may be trapped in a local minimum. 
In this paper, the premature convergence judgement threshold $n_{\rm{th}}$ is fixed at 5.
In the modified PSO algorithm, we add a perturbation to some elements of $\bm{P}_{gbest}^{n}$ in order to save the swarm from local optima. 
Particularly, we randomly choose the $m$th element of the best position record of the swarm, i.e. $\bm{P}_{gbest}^{n}(m)$, which satisfied ${{{\bf{w}}_{gbest}}(m) = 1}$. Then, we change $\bm{P}_{gbest}^{n}(m)$ into $\tilde {\bm{P}}_{gbest}^n(m)$ by $\widetilde P_{gbest}^n(m){\rm{ = (1}} - {\rm{0}}{\rm{.1}}{r_{\rm{3}}}{\rm{)}}P_{gbest}^n(m)$, where $r_3$ is a random number uniformly distributed in the range $\left[ {0,1} \right]$.
The pseudocode of the modified PSO algorithm is shown in Algorithm 1, where $j$ is omitted for the sake of simplicity. In addition, the computational complexity of the modified PSO is ${\mathcal{O}}(n_{\rm{max}}\lambda_{\rm{max}})$.
\begin{table}
\begin{center}
\begin{tabular}{l}
\hline\noalign{\smallskip}
\textbf{Algorithm 1:} The Modified PSO Algorithm\\
\noalign{\smallskip}\hline
1:  \;\textbf{Input:} Swarm size $\lambda_{\rm{max}}$, maximum iterations $n_{\rm{max}}$, premature convergence judgement threshold $n_{\rm{th}}$; \\
2:  \;\textbf{Random initialization:} \\
 \;\quad Position of each particle: $\bm{P}_\lambda= \left[ {{{\bf{ w }}_\lambda },{\alpha _\lambda }} \right], 1 \le \lambda \le \lambda _{\rm{max}} $;\\
 \;\quad Velocity of non-integer variables of each particle: ${\bm{V}_\lambda }, 1 \le \lambda \le \lambda _{\rm{max}}$;\\
 \;\quad Particle's best known position: 
$\bm{P}_{\lambda,pbest} = \left[ {\bf{w }}_{\lambda,pbest},{\alpha _{\lambda,pbest}} \right], 1 \le \lambda \le \lambda _{\rm{max}}$; \\
 \;\quad Swarm's best known position: $\bm{P}_{gbest}= \left[ {{{\bf{ w }}_{gbest}},{\alpha _{gbest}}} \right]$; \\
3:  \;\textbf{While} $n \le n_{\rm{max}}$\\
4:  \;\quad \textbf{For} each particle \\
5:  \;\quad \quad Update the velocity and the position of each particle according to (\ref{equ-PSO-velocity}) and (\ref{equ-PSO-position}); \\
6:  \;\quad \quad Calculate the fitness value of each particle by (\ref{op-DPP-intracell-objective});\\
7:  \;\quad \quad \textbf{If} the fitness value of ${\bm{P}}_\lambda^{n}$ is smaller than the fitness value of ${\bm{P}}_{\lambda,pbest}^{n-1}$ \\
8:  \;\quad \quad \quad Update the particle's best known position as ${\bm{P}}_{\lambda,pbest}^{n}={\bm{P}}_\lambda^{n}$;\\
9: \quad \quad \textbf{End if} \\
10: \quad \quad \textbf{If} the fitness value of ${\bm{P}}_\lambda^{n}$ is smaller than the fitness value of ${\bm{P}}_{gbest}^{n-1}$\\
11: \quad \quad \quad Update the swarm's best known position as ${\bm{P}}_{gbest}^{n}={\bm{P}}_\lambda^{n}$; \\
12: \quad \quad \textbf{End if} \\
13: \quad \textbf{End for}\\
14: \quad \textbf{If} ${\bm{P}}_{gbest}^{n}={\bm{P}}_{gbest}^{n-n_{\rm{th}}}$\\
15:  \;\quad \quad Change ${\bm{P}}_{gbest}^{n}$ into $\tilde {\bm{P}}_{gbest}^n$ by adding a perturbation;\\
16:  \;\quad \quad Calculate the fitness value of $\tilde {\bm{P}}_{gbest}^n$ by (\ref{op-DPP-intracell-objective});\\
17:  \;\quad \quad \textbf{If} the fitness value of $\tilde {\bm{P}}_{gbest}^n$ is smaller than the fitness value of ${\bm{P}}_{gbest}^{n}$;\\
18: \quad \quad \quad Update the swarm's best known position as ${\bm{P}}_{gbest}^{n}=\tilde{\bm{P}}_{gbest}^n$;\\
19:  \;\quad \quad \textbf{End If}\\
20: \quad \textbf{End If}\\
21: \textbf{End while}\\
22: \textbf{Output:} ${{\bf{w}}^*}$ and ${\alpha}^{*}$ in the swarm's best known position ${\bm{P}}_{gbest}^{n_{\rm{max}}}$. \\
\noalign{\smallskip}\hline
\end{tabular}
\end{center}
\vspace*{-8mm}
\end{table}

\subsubsection{IG-based greedy user-centric scheduling}
\label{subsection-IG}
After solving a series of inner problems and finding the secure parameters of each user, we formulate the user-centric scheduling as an outer problem, which is expressed as
\begin{subequations}
\label{op-outer}
\begin{alignat}{2}
&\!\min_{\bf{\Pi }} &\;\;&  \sum\limits_{j = 1}^{{K_u}} {\left\{ {\frac{1}{{{\theta _j}}}\log \left[ {{e^{ - {\theta _j}\tilde R_j^{\rm{s}}({\alpha _j},{{\bf{w}}_j},{{\bf{\Pi }}_j})}}} \right] + {F_j}\left[ {{e^{ - {\theta _j}\tilde R_j^{\rm{s}}({\alpha _j},{{\bf{w}}_j},{{\bf{\Pi }}_j})}} - {e^{ - B_j^{\rm{e}}{\theta _j}}}} \right]} \right\}}  \label{outer-OP-obj}\\
& \;\;\text{s.t.} &      & {\pi _{i,j}}(t) \in \{ 0,1\},\label{outer-OP-con1} \\
&                  &      & \sum\limits_{i = 1}^{{K_a}} {{\pi _{i,j}}(t)} \sum\limits_{i = 1}^{{K_a}} {{\pi _{i,k}}(t)}  = 0,\;\;j,k \in \left\{ {1, \ldots ,{K_u}|{\Omega _j} \cap {\Omega _k} \ne \emptyset } \right\}.\label{outer-OP-con2}
\end{alignat}
\end{subequations}

To solve problem (\ref{op-outer}) and find the combination of simultaneous transmission secure cells, we propose an IG-based greedy user-centric scheduling algorithm. 
To eliminate the interference between users and satisfy constraint (\ref{outer-OP-con2}), we first define the IG to characterize the interference relationship between users. 
According to the layout of APs and users, the IG is generated as an undirected graph $\mathcal{G}(\mathcal{V};\mathcal{E})$. The vertex set $\mathcal{V}(\mathcal{G})$ of the IG is the user set $\Phi$ and the vertex $v_j$ corresponds to the user $j$. The edge set $\mathcal{E}(\mathcal{G})$ of the IG is determined by the interuser interference.
Specifically, an edge between vertex $v_j$ and vertex $v_k$ exists when their AP sets satisfy $\Omega_j \cap {\Omega _k} \ne \emptyset$. According to graph theory, two adjacent vertices connected by an edge are identified as neighbours, which means two corresponding users may interfere each other. The set of vertex $v_j$'s neighbours is denoted by $\mathcal{N}(v_j)$. 
Note that the channel gain threshold $\varepsilon$ of the capable AP set $\Omega_j$ has a significant impact on IG. 
According to the definition of $\Omega_j$ in (\ref{capable AP set}), if $\varepsilon$ is too small, all the weak interference links would be taken into account in IG, which leads to too much intricate edges in IG.
To build an interference-free scheduling scheme, we suppose that neighbour users cannot be scheduled at the same time. 
Therefore, in this case, some users cannot be scheduled for a long time, especially the users located at the edge of the room.
In contrast, if $\varepsilon$ is too large, a high interference caused by the links whose channel gains are less than $\varepsilon$ is ignored in the scheduling process, which leads to inaccurate scheduling results.
In this paper, to set $\varepsilon$, we divide the receiving plane into a number of positions. For each position, we calculate the sum of NLoS channel gain and set $\varepsilon$ as the sum value that can be achieved by more than $90\%$ of the positions.
Fig. \ref{Fig-6subplot}a and Fig. \ref{Fig-6subplot}d show two examples of IG embedded in the top view of a multi-user multi-AP VLC system, in which solid lines represent the edges of the IG.

Based on the IG, we propose a greedy user-centric scheduling algorithm. 
First, the initial cell formation matrix ${\bf{\Pi '}}$ is determined according to the AP sets $\Omega_j, j=1,...,K_u$. More specifically, we initialize ${\pi '_{i,j}} = 1$ if $i \in \Omega_j$, and initialize ${\pi '_{i,j}} = 0$, otherwise. The dotted lines in Fig. \ref{Fig-6subplot}a and Fig. \ref{Fig-6subplot}d show the connections between APs and users according to ${\bf{\Pi'}}$.
Further, with the obtained ${\alpha}_j$ and ${\bf{w}}_j$ from problem (\ref{op-DPP-intracell}), we calculate the DPP value of user $j$ by (\ref{op-DPP-intracell-objective}). 
Then, by treating the DPP value of each user as the weight of each vertex, we transform the IG into a weighted IG. The user-centric scheduling is formulated as a minimum-weighted independent-set problem of the weighted IG. According to the min-greedy strategy, for each scheduling, we extract the vertex $v_j$ with minimum DPP value and add it into the independent-set $\mathcal{I}$ and then delete it and its neighbours $\mathcal{N}(v_j)$ from the original vertex set $\mathcal{V}(\mathcal{G})$. The `extract and delete' process is repeated until no vertex is in $\mathcal{V}(\mathcal{G})$. According to the independent set $\mathcal{I}$, we can obtain the secure cell formation matrix ${\bm{\Pi}}^*$. In detail, we have $\pi_{i,j}=1$ if $v_j \in \mathcal{I}, i \in \Omega_j$ and  $\pi_{i,j}=0$, otherwise. The detailed process of the proposed IG-based greedy user-centric scheduling is illustrated in Algorithm 2. In addition, the computational complexity of Algorithm 2 is ${\mathcal{O}}(K_u)$.
\begin{table}
\begin{center}

}
\vspace*{-16mm}
\caption{(a)(d) The top view of multi-AP multi-user VLC system with different kinds of users' location. There are 10 users and $8 \times 8$ APs. FoV is $\varphi_{\rm{c}}=100^\circ$. (b)(c) The user-centric secure cell formation results in different timeslots. Users' location is shown as Fig. 4a. (e)(f) The user-centric secure cell formation results in different timeslots. Users whose location is shown as Fig. 4d.}
\label{Fig-6subplot}
\end{figure}
\section{Numerical and Simulations Results}
\label{section-simulation}
In this section, we present the simulation results to evaluate the proposed delay driven secure cell formation algorithm. We consider a three-dimensional space with size of 16 m $\times$ 16 m $\times$ 2.5 m. Optical APs are uniformly installed at the ceiling of the room. Users are randomly distributed in the room and the height of the receiving plane is 0.5 m. Both LoS links and the first reflected links are taken into consideration. To calculate the channel gains of the reflected links, the surfaces of four walls are divided into a number of small areas (0.1 m $\times$ 0.05 m). By summing the received power from the reflection of these small areas, we get the channel gains of the first reflected links. 
Besides, we assume that all users have the same FoV. Each user has its own statistical delay QoS requirement that is characterized by the QoS exponent. A user with a large index has a strict QoS requirement. In Fig.~\ref{Fig-PSO-conver}-Fig.~\ref{Fig-SER-effects}, users' QoS exponents are set to increase with users' index, uniformly from ${\rm{10}}^{\rm{ - 10}}$ to ${\rm{10}}^{\rm{ - 7}}$. In addition, users' effective bandwidth is also set to increase with users' index, uniformly from ${\rm{10}}^{\rm{ 5}}$ to ${\rm{10}}^{\rm{ 6}}$ in Fig.~\ref{Fig-PSO-conver}-Fig.~\ref{Fig-SER-effects}. We evaluate the performance of the proposed secure cell formation algorithm on both short-term and long-term. In addition, we also analyse the effect of various factors on the ESR. The main system parameters are summarized in Table I.

\begin{table}[htbp]
\caption{Simulation parameters}
\begin{center}
\begin{tabular}{ll}      
\hline\noalign{\smallskip}
\textbf{Name of Parameter} & \textbf{Value of Parameter}   \\
\noalign{\smallskip}\hline\noalign{\smallskip}
Available bandwidth, $B_a$ & 20 MHz\\
The number of LEDs per AP, $m$ & 400\\
DC offset current, $I_{DC}$ & 700 mA\\
Modulation index, $\gamma$ & 0.2\\
Refractive index of lens at a PD, $a$  & 1.5  \\
Physical area in a PD, $\delta$   & $10^{-4} \;\rm{m}^2$\\
Semi-angle at half power, ${\phi _{\;1/2}}$ & 70 deg. \\
Gain of the transimpedance amplifier, $T$  & 1 V/A\\
Current-to-light conversion efficiency, $\eta$ & 0.44 W/A\\
Responsivity of a PD, $\varpi$ & 0.54 A/W\\
Reflectance factor, $\rho$ & 0.8 \\
\noalign{\smallskip}\hline
\end{tabular}
\label{Table-parameters}
\end{center}
\end{table}
\subsection{Results of the per-slot drift-plus-penalty problem}

Different kinds of users' locations generated by uniform distribution are considered in our simulation. Fig.~\ref{Fig-6subplot}a and Fig.~\ref{Fig-6subplot}d show the top view of the multi-AP multi-user system with two different kinds of users' locations. The dotted lines represent the connections according to the initial cell formation matrix $\bm{\Pi}'$ and the solid lines represent the edges in the interference graph. Fig.~\ref{Fig-PSO-conver} substantiates the convergence of the proposed  modified PSO algorithm. Since PSO is one of the population-based intelligent algorithms, enlarging the size of the swarm would increase the diversity of particles' position and expand the search scope of feasible solutions. A large swarm size may accelerate the convergence of modified PSO algorithm, but also aggravate the computational complexity at the same time. 
In the following simulations, we choose swarm size $\lambda _{\rm{max}}=40$ and the maximum iterations ${n_{\rm{max}}}=100$. 

By employing the modified PSO algorithm and IG-based greedy user-centric scheduling algorithm, we can obtain the user-centric secure cell formation results as shown in Fig.~\ref{Fig-6subplot}b-c and Fig.~\ref{Fig-6subplot}e-f, which correspond to the users' locations in Fig.~\ref{Fig-6subplot}a and Fig.~\ref{Fig-6subplot}d, respectively. In each secure cell, artificial noise with specified power separation factor is adopted to enhance the transmission security. Comparing to the initial cell formation $\bm{\Pi}'$, we can see that the users who may interfere each other are scheduled in different timeslots according to the DPP value calculated by (\ref{op-DPP-st-objective}).
\definecolor{mycolor1}{rgb}{0.46667,0.67451,0.18824}%
\definecolor{mycolor2}{rgb}{0.30196,0.74510,0.93333}%
\definecolor{mycolor3}{rgb}{0.85098,0.32549,0.09804}%
\subsection{Results of long-term statistical delay guaranteed secure cell formation problem}
By tackling short-term DPP problems slot by slot, the long-term performance of the proposed secure cell formation algorithm is evaluated in this subsection. Taking the users' locations as Fig.~\ref{Fig-6subplot}a for example, we show  cumulative distribution function (CDF) of the average achievable secrecy rate per user of various timeslots in Fig. \ref{Fig-pri-rate-slot-peruser}. The relationship of the results of successive timeslots depends on the evolution of Lyapunov virtual queues. We compare our proposed Lyapunov optimization based cell formation algorithm with the following scheduling algorithms \cite{Chi-2018-COPP}.

\begin{itemize}
\item Maximum Rate (MR): Based on the IG, the achievable rate without artificial noise of each user is treated as the weight of each vertex. The achievable rate without artificial noise of user $j$ is expressed as \cite{Moser-MISO}
\begin{equation}
{r_j}(t) = \frac{1}{2}\log \left[ {{\rm{1 + }}\frac{{4{{\bf{h}}_j^T(t)}{{\bf{h}}_j(t)}{A^2}}}{{2\pi e{\sigma ^2}}}} \right].\nonumber
\end{equation}
User scheduling is modelled as a maximum-weighted independent-set problem. According to the greedy algorithm, if there are edges among several users, only the user with maximum achievable rate among its neighbours can be scheduled.
\item Proportional Fairness (PF): Based on the IG, we define the weight of each vertex by the scheduling priority and tackle the maximum-weighted independent-set problem by greedy algorithm. The scheduling priority of user $j$ is expressed as $\kappa _j{(t)} = {{r_j{(t)}} \mathord{\left/
 {\vphantom {{r_j{(t)}} {C_j{(t)}}}} \right.
 \kern-\nulldelimiterspace} {C_j{(t)}}},\;j = 1, \ldots ,{K_u}$. Parameter ${C_{j}(t)}$ represents the average achievable rate of user $j$ in timeslot $t$. To allow all users at least a minimal level of service, ${C_j(t)}$ is updated as follows
\begin{equation}
C_j{(t)} = \left\{
 \begin{aligned}&(1 - \frac{1}{{{T_F}}})C_j{\left( {t - 1} \right)} + \frac{1}{{{T_F}}}r_j{\left( {t - 1} \right)},
\;\;{\rm{if\;user }}\;j{\rm{\;is\;scheduled\;in\;timeslot\;}}t,\\
&(1 - \frac{1}{{{T_F}}})C_j{\left( {t - 1} \right)},
\;\;{\rm{if\;user }}\;j{\rm{\;is\;not\;scheduled\;in\;timeslot\;}}t,
\end{aligned} \right. \nonumber
\end{equation}
where ${T_F}$ is constant window size.
\item MR with artificial noise: We implement the artificial noise approach in each cell formed by MR method.
\item PF with artificial noise: We implement the artificial noise approach in each cell formed by PF method.
\end{itemize} 

The power separation factors of the compared algorithms are fixed and set as 0.7 in Fig. \ref{Fig-pri-rate-slot-peruser}. Fig. \ref{Fig-pri-rate-slot-peruser} shows that the proposed secure cell formation algorithm is capable of providing higher average secrecy rate than MR and PF approaches. There are two main reasons of the improvement. One of them is that the proposed cell formation provides the customized precoding and intensity of artificial noise in each secure cell. Another is that the proposed cell formation algorithm schedules users with larger DPP values, which take users' security into account.

\definecolor{mycolor1}{rgb}{1.00000,0.00000,1.00000}%
\definecolor{mycolor2}{rgb}{0.00000,1.00000,1.00000}%

\begin{figure}[!t]
\centering
\begin{minipage}[t]{0.48\linewidth}
\resizebox{3in}{!}{%
%
}
\vspace*{-6mm}
\caption{ (a) Convergence performance of PSO algorithm with different swarm size for the user in Fig.~\ref{Fig-6subplot}a. (b) Convergence performance of PSO algorithm with different swarm size for the user in Fig.~\ref{Fig-6subplot}d.}
\label{Fig-PSO-conver}
\end{minipage}
\hfill
\begin{minipage}[t]{0.48\linewidth}
\resizebox{3in}{!}{%
%
%
}
\vspace*{-6mm}
\caption{CDF of the average achievable secrecy rate per user of different cell formation methods, where the number of timeslots is 150, unblocking probability is $\beta=0.7$ and FoV is $\varphi_{\rm{c}}=100^\circ$.}
\label{Fig-pri-rate-slot-peruser}
\end{minipage}
\end{figure}

Fig.~\ref{Fig-vir-queue-subplot} shows the change of backlog of Lyapunov virtual queues for users with different QoS exponents. We can see that the backlog of all users does not tend to infinity, in other words, all the Lyapunov virtual queues are stabilized. Therefore, according to Lemma 1, all users' statistical delay constraints can be satisfied in the long-term. Besides, the fluctuation of the backlog of virtual queues becomes larger when the QoS exponent becomes bigger. This is reasonable because a bigger QoS exponent corresponds to a stricter delay QoS requirement, which is more difficult to guarantee. 

The average ESR per user of different cell formation methods is shown in Fig.~\ref{Fig-averEC}. The ESR of each user is calculated according to formula (\ref{EC-limit}). Different from the achievable secrecy rate in Fig.~\ref{Fig-pri-rate-slot-peruser}, ESR is a long-term metric, which gradually converges over timeslots. 
Comparing to the MR and PF scheduling methods, our proposed secure cell formation algorithm achieves higher delay-constrained secrecy rate. The reason is that our user-centric secure cell formation is a result of a joint optimization considering both users' delay and security guarantee when the user scheduling is decided based on users' DPP values.

\begin{figure}[!t]
\centering
\begin{minipage}[t]{0.48\linewidth}
\pgfplotsset{scaled y ticks=false}
\resizebox{3in}{!}{%
%
}
\vspace*{-6mm}
\caption{Fluctuation of Lyapunov virtual queue of each user, where users' QoS exponents are set to increase with users' index, uniformly from ${\rm{10}}^{\rm{ - 10}}$ to ${\rm{10}}^{\rm{ - 7}}$.}
\label{Fig-vir-queue-subplot}
\end{minipage}
\hfill
\begin{minipage}[t]{0.48\linewidth}
\resizebox{3in}{!}{%
%
%
}
\vspace*{-6mm}
\caption{Average effective secrecy rate of different cell formation algorithms, where unblocking probability is $\beta=0.7$ and FoV is $\varphi_{\rm{c}}=100^\circ$.}
\label{Fig-averEC}
\end{minipage}
\vspace{-4mm}
\end{figure}

In Fig.~\ref{Fig-SER-effects}, we analyse the effects of the FoV and the number of random located users on the ESR. For each scenario, 10 kinds of users' locations are generated randomly and the average ESR is obtained in Fig.~\ref{Fig-SER-effects}. As expected, our proposed secure cell formation algorithm remains superior in all scenarios considered. Fig.~\ref{Fig-averEC-fov} shows that the ESR per user decreases when the FoV increases. This is reasonable because a large FoV at receiver can accept more LoS links, which increases the possibility of interference among users. Besides, as shown in Fig.~\ref{Fig-averEC-num}, when the number of users is increased, the ESR per user is reduced, due to the increased resource competition. 
\vspace*{6mm}

\begin{figure}[!t]
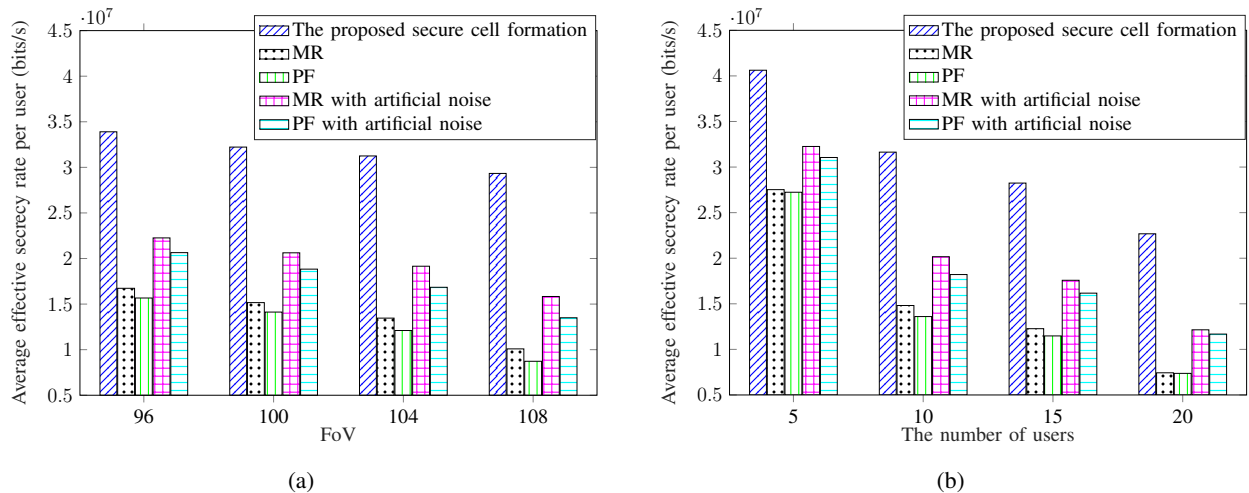

\centering
\subfloat[\label{Fig-averEC-fov}]{%
\resizebox{3.1in}{!}{%
%
}}

\caption{(a) Average effective secrecy rate for various FoV, where the number of users is 10 and unblocking probability is $\beta=0.7$. (b) Average effective secrecy rate for various number of users, where unblocking probability is $\beta=0.7$ and FoV is $\varphi_{\rm{c}}=100^\circ$.}

\label{Fig-SER-effects}
\vspace{-4mm}
\end{figure}

\noindent
\begin{minipage}{\textwidth}
\makeatletter\def\@captype{figure}\makeatother
\begin{minipage}[b]{0.5\linewidth}
\centering
\resizebox{3.3in}{!}{%
%
%
}
\vspace*{-1.2cm}
\caption{Normalized average backlog of Lyapunov virtual queues for various $D_u$ and $B^{\rm{e}}$, where $\theta=10^{-8}$ for each user.}
\label{Fig-maxnum}
\end{minipage}
\hfill
\makeatletter\def\@captype{table}\makeatother
\begin{minipage}[b]{0.5\linewidth}
\caption{The ADF test result for Normalized average backlog of Lyapunov virtual queues in Fig. 10}
\label{table-ADF}
\centering
\resizebox{!}{1.5in}{%
\begin{tabular}[b]{ |c|c|c|c| } 
 \hline
 \diagbox{\footnotesize{$B^{\rm{e}}$}}{\footnotesize{$D_u$}}& \footnotesize{$1/9$} & \footnotesize{$1/12$} & \footnotesize{$1/16$} \\ 
   \hline
\footnotesize{$10^6$} & \footnotesize{0} & \footnotesize{1} & \footnotesize{1}  \\ 
 \hline
\footnotesize{$5\times10^6$} & \footnotesize{0} & \footnotesize{0} & \footnotesize{1} \\ 
 \hline
 \footnotesize{$10^7$} & \footnotesize{0} & \footnotesize{0} & \footnotesize{0} \\ 
 \hline
\end{tabular}
}

\end{minipage}
\end{minipage}

\vspace*{6mm}
To show the system capability in terms of the user density $D_u (\rm{users}/\rm{m}^2)$ and the QoS requirement, we provide Fig.~\ref{Fig-maxnum}, in which we use the stability of Lyapunov virtual queues to characterize whether the system works well or not. 
In addition, the augmented Dickey-Fuller test (ADF) is used to analytically judge each time sequence in Fig.~\ref{Fig-maxnum} is stability or not \cite{ADF}. 
Particularly, the ADF test result equal to 1 means the time sequence is stable and vice versa \cite{ADF-Matlab}.
As shown in Fig.~\ref{Fig-maxnum}, for ${B^{\rm{e}}}=10^6 \;({\rm{bits}/\rm{s}}), D_u=1/12\;(\rm{users}/\rm{m}^2)$, ${B^{\rm{e}}}=10^6\; ({\rm{bits}/\rm{s}}), D_u=1/16\; ({\rm{users}/\rm{m}^2})$ and ${B^{\rm{e}}}=5\times10^6 \;({\rm{bits}/\rm{s}}), D_u=1/16 \;(\rm{users}/\rm{m}^2)$, the Lyapunov virtual queues are stable and the corresponding ADF test results in Table II are equal to 1, which means the system can work normally. 
With the increase of ${B^{\rm{e}}}$ and $D_u$, the virtual queues tend to be unstable and the backlog of the virtual queue would go to infinity over time.

\section{Conclusion}
\label{section-conclusion}
In this paper, we proposed a user-centric secure cell formation algorithm with heterogeneous statistical QoS guarantees for indoor VLC networks. Different intensities of artificial noise are adopted in the user-centric secure cells, which helps to enhance the secrecy performance. 
Based on the EC theory, ESR is derived to evaluate the transmission rate constrained by both secrecy and statistical delay requirements of each user. Further, users' probabilistic delay guarantee goals are modelled as long-term constraints and the user-centric secure cell formation problem is formulated as a long-term optimization problem. Thanks to Lyapunov optimization theory, we decomposed this problem into several tractable per-slot problems and solved them by an modified PSO and IG-based scheduling algorithm.  
The proposed Lyapunov optimization-based statistical delay QoS guarantee method may stimulate a range of further research when delay needs to satisfy a probabilistic constraint.
Moreover, the concept of ESR may also inspire some further research by providing a fresh perspective for wireless resource allocation with both secrecy and delay guarantee.


%

\appendices
\section{Proof of Theorem 1}
\label{Appendix-A}
We follow the approach proposed in \cite{Lampe-2015-JSAC} to derive a lower bound of achievable secrecy rate. For the sake of simplicity, the timeslot variable $t$ is omitted from our notations in this proof.

According to the results of \cite{Anas-2019-TCOM} and \cite{1978-TIT}, for the wiretap channel given by (\ref{recev-signal-AN}) and (\ref{recev-signal-AN-eve}), the lower bound of achievable secrecy rate with artificial noise aided between user $j$ and user $k$ can be given by 

\begin{equation}
\label{formula-A1}
\begin{aligned}
R_{j,k}^{{\rm{s}}} &\ge {\left[ {I({{\bf{w}}_j}{s_j^u};{y_j}) - I({{\bf{w}}_j}{s_j^u};{y_{j,k}^{\rm{e}}})} \right]^ + }\\
& = {\left[ {h({y_j}) - h(\left. {{y_j}} \right|{{\bf{w}}_j}{s_j^u}) - h(y_{j,k}^{{\rm{e}}}) + h(\left. {y_{j,k}^{{\rm{e}}}} \right|{{\bf{w}}_j}{s_j^u})} \right]^+, }
\end{aligned}
\end{equation}
where $I( \cdot \;;\; \cdot )$ denotes the mutual information, $h( \cdot )$ denotes the differential entropy and $h( \cdot \;|\; \cdot )$ denotes the conditional entropy. 

We assume the useful signal $s_j^u$ follows the uniform distribution $\left[ { - {A_j^u},{A_j^u}} \right]$ and the jamming signal $s_{j_l}^a$ follows the uniform distribution $\left[ { - {A_j^a},{A_j^a}} \right]$. According to the relationship between the transmitted power and the signal distribution, we have ${A_j^u}=\alpha_j A$  and ${A_j^a}=\frac{{(1 - {\alpha _j})A}}{{\left| {{\Omega _j}} \right| - 1}}$.

Then, we derive each term of formula (\ref{formula-A1}). First, recalling the formula (\ref{received-signal-eve}) and using the entropy-power inequality (EPI) \cite{EPI}, we get the lower bound of $h({y_j})$ as
\begin{equation}
\label{formula-A2}
\begin{aligned}
h({y_j})& = h({\bf{h}}_j^T{{\bf{w}}_j}{{s_j^u}}{\rm{ + }}{n_j})\\
&\mathop  \ge \limits^{} \frac{1}{2}\log (\exp (2h({\bf{h}}_j^T{{\bf{w}}_j}s_j^u)) + \exp (2h({n_j})))\\
& = \frac{1}{2}\log (4{\bf{w}}_j^T{\bf{h}}_j^{}{\bf{h}}_j^T{{\bf{w}}_j}\left({A_j^u}\right)^2 + 2\pi e\sigma _j^2).
\end{aligned}
\end{equation}

Similarly, we can derive the conditional entropy $h(\left. {y_{j,k}^{{\rm{e}}}} \right|{{\bf{w}}_j}{s_j^u})$ as
\begin{equation}
\label{formula-A3}
\begin{aligned}
h(\left. {y_{j,k}^{{\rm{e}}}} \right|{{\bf{w}}_j}{s_j^u})
& = h\left( {{{\left( {{\bf{h}}_{j,k}^{\rm{e}}} \right)}^T}{\sum\limits_{l = 1}^{\left| {{\Omega _j}} \right| - 1} {{{\hat \Gamma }_{{j_l}}}s_{{j_l}}^a} }{\rm{ + }}n_{j,k}^{\rm{e}}} \right)\\
& \ge \frac{1}{2}\log \left[ {\exp \left( {2 \cdot \frac{1}{2}\log \left( {\sum\limits_{l = 1}^{\left| {{\Omega _j}} \right| - 1} {\exp \left( {2h\left( {{{\left( {{\bf{h}}_{j,k}^{\rm{e}}} \right)}^T}{{\hat \Gamma }_{{j_l}}}s_{{j_l}}^a} \right)} \right)} } \right)} \right) + \exp \left( {2h(n_{j,k}^{\rm{e}})} \right)} \right]\\
& = \frac{1}{2}\log \left[ {\sum\limits_{l = 1}^{\left| {{\Omega _j}} \right| - 1} {4\hat \Gamma _{{j_l}}^T{\bf{h}}_{j,k}^{\rm{e}}{{\left( {{\bf{h}}_{j,k}^{\rm{e}}} \right)}^T}{{\hat \Gamma }_{{j_l}}}{{\left( {A_j^a} \right)}^2}}  + 2\pi e\left( {\sigma _{j,k}^{\rm{e}}} \right)_{}^2} \right].
\end{aligned}
\end{equation}

Second, since the VLC noise follows a Gaussian distribution, we have
\begin{equation}
\label{formula-A4}
h(\left. {{y_j}} \right|{{\bf{w}}_j}{s_j^u}) = h({n_j}) = \frac{1}{2}\log (2\pi e\sigma _j^2).
\end{equation}

Third, an upper bound on $h(y_{j,k}^{{\rm{e}}})$ can be obtained using the differential entropy of a Gaussian random variable with variance ${\mathbb{V}\rm{ar}} (y_{j,k}^{{\rm{e}}})$. Therefore, we can obtain
\begin{equation}
\label{formula-A5}
\begin{aligned}
&h(y_{j,k}^{{\rm{e}}}) \le \frac{1}{2}\log \left[ {2\pi e{\mathbb{V}\rm{ar}} \left( {{{\left( {{\bf{h}}_{j,k}^{\rm{e}}} \right)}^T}{{\bf{w}}_j}s_j^u} \right.} \right.\left. {\left. { + {{\left( {{\bf{h}}_{j,k}^{\rm{e}}} \right)}^T}\sum\limits_{l = 1}^{\left| {{\Omega _j}} \right| - 1} {{{\hat \Gamma }_{{j_l}}}s_{{j_l}}^a}  + n_{j,k}^{\rm{e}}} \right)} \right]\\
&=\frac{1}{2}\log \left[ {\frac{{2\pi e}}{3}\left( {{\bf{w}}_j^T{\bf{h}}_{j,k}^{\rm{e}}{{\left( {{\bf{h}}_{j,k}^{\rm{e}}} \right)}^T}{{\bf{w}}_j}{{\left( {A_j^u} \right)}^2}} \right.} \right.\left. {\left. { + \sum\limits_{l = 1}^{\left| {{\Omega _j}} \right| - 1} {\hat \Gamma _{{j_l}}^T{\bf{h}}_{j,k}^{\rm{e}}{{\left( {{\bf{h}}_{j,k}^{\rm{e}}} \right)}^T}{{\hat \Gamma }_{{j_l}}}{{\left( {A_j^a} \right)}^2}}  + 3\left( {\sigma _{j,k}^{\rm{e}}} \right)_{}^2} \right)} \right].
\end{aligned}
\end{equation}

By substituting the expressions (\ref{formula-A2})-(\ref{formula-A5}) in (\ref{formula-A1}), we get Theorem 1.

\ifCLASSOPTIONcaptionsoff
  \newpage
\fi



%



\bibliographystyle{IEEEtran}
\bibliography{IEEEabrv,references-Lei-paper-Lyapunov}
%

%
%
%
%
%
%
%
%
%
%
%




\end{document}